\documentclass[11pt,a4paper]{article}

\usepackage[english]{babel}
\usepackage[latin1]{inputenc}
\usepackage[T1]{fontenc}
\usepackage{graphicx}
\usepackage{xcolor}
\usepackage{amsmath}
\usepackage{amssymb}
\usepackage[final]{pdfpages}
\usepackage{graphicx}
\usepackage{bbold}
\usepackage{dsfont}
\usepackage[tight,italian]{minitoc}
\usepackage{caption}
\usepackage{subcaption}
\usepackage{subeqnarray}
\usepackage{multirow}
\usepackage{multicol}
\usepackage[toc,page]{appendix}
\usepackage[normalem]{ulem}
\usepackage{booktabs}
\usepackage[font=small,labelfont=bf]{caption}
\usepackage[bitstream-charter, greekfamily=default]{mathdesign}

\usepackage[scale=0.8]{geometry}

\newenvironment{changemargin}[2]{%
\begin{list}{}{%
\setlength{\topsep}{0pt}%
\setlength{\leftmargin}{#1}%
\setlength{\rightmargin}{#2}%
\setlength{\listparindent}{\parindent}%
\setlength{\itemindent}{\parindent}%
\setlength{\parsep}{\parskip}%
}%
\item[]}{\end{list}}
 
\title{How does latent liquidity get revealed in the limit order book?}

\usepackage{authblk}

\author[1]{Lorenzo Dall'Amico{$^*$}}
\author[1,2]{Antoine Fosset\footnote{Both authors contributed equally to this work.}}
\author[2]{Jean-Philippe Bouchaud}
\author[1,2]{Michael Benzaquen\footnote{Corresponding author: michael.benzaquen@polytechnique.edu}}
\affil[1]{Ladhyx UMR CNRS 7646, Ecole polytechnique, 91128 Palaiseau Cedex, France}
\affil[2]{Capital Fund Management, 23 rue de l'Universit\'e, 75007, Paris, France}
\date{}
\begin{document}

\date{\today}
\maketitle

\abstract{
Latent order book models have allowed for significant progress in our understanding of price formation in financial markets.  In particular they are able to reproduce a number of stylized facts, such as the square-root impact law. An important question that is raised -- if one is to bring such models closer to real market data -- is that of the connection between the latent (unobservable) order book and the real (observable) order book. Here we  suggest a simple, consistent mechanism for the revelation of latent liquidity that allows for quantitative estimation of the latent order book from real market data. We successfully confront our results to real order book data for over a hundred assets and discuss market stability. One of our key theoretical results is the existence of a market instability threshold, where the conversion of latent order becomes too slow, inducing \emph{liquidity crises}. Finally we compute the price impact of a metaorder in different parameter regimes.   
}

\vspace{0.5cm}

\section*{Introduction}

In the past few years, price formation in financial markets has attracted the interest of a broad community of academics and practitioners. {This can partially be explained by the growing quality of market data which now allows to test theories with levels of precision that have nothing to envy to natural sciences, see e.g. \cite{kanazawa2018derivation}.} The study of market impact is both of fundamental and practical relevance. Indeed, while it is an important component for understanding price dynamics, it is also the source of substantial trading costs. Despite a few dissenting voices, the empirical nonlinear  impact of metaorders (often coined  the \emph{square root law}) is now indisputably among the most robust stylized facts of modern finance (see e.g. \cite{Toth2011,Torre1997,Almgren2005,engle2006measuring,mastromatteo2014agent,Brockmann2015,bacry2015market,bershova2013non}). 
 In particular, for a given asset the expected average price return $I$ between  the beginning and the end of a metaorder of size $Q$ follows $I(Q)=Y\sigma_\textrm{d} (Q/V_\textrm{d})^\theta$ where $Y$ is a numerical factor of order one,  $\sigma_\textrm{d}$ and $V_\textrm{d}$ denote daily volatility and daily traded volume respectively, and $\theta <1$ (typically $0.4<\theta<0.6$) bears witness of the concave nature of price impact.\\

Recently, a new class of agent based models (ABM) has helped to gain insight into  the origins of the square root law. Given that the instantaneous liquidity revealed in the limit order book is very small (less than $1\%$ of daily traded volume), \emph{latent} order book models \cite{Toth2011,mastromatteo2014agent,lehalle2011high} 
  build on the idea that revealed liquidity chiefly reflects the activity of high frequency market makers that act as intermediaries between much larger \emph{unrevealed} volume imbalances. The latter are not revealed in the limit order book in order to avoid giving away precious private information, until the probability to get executed is large enough to warrant posting the order close to the bid (or to the ask). Within this class of ABM, reaction-diffusion models (see e.g. \cite{bak1997price,donierLLOB,benzaquen2018market}) 
   have proved very successful at reproducing the square root law in a setup free of price manipulation, and more recently consistent with the so-called diffusivity puzzle -- that is, reconciling persistent order flow with diffusive price dynamics \cite{benzaquen2018market,benzaquen2018fractional}. However, one very important question is yet to be addressed if one is to connect such models with real observable and quantifiable data: what is the relation between revealed and latent liquidity and what are the mechanisms through which latent liquidity becomes revealed? These issues are the subject of the present communication. We propose a simple dynamic model to account for liquidity flow between the latent and revealed order books. We introduce two ingredients that are to our eyes essential to reproduce realistic limit order book shapes: (i) the incentive to reveal one's liquidity increases with decreasing distance to the trade price, and (ii) the process of revealing latent liquidity is not instantaneous and lag effects may be an important source of instability, as real liquidity is found to vanish in certain regions of parameters. In addition to providing an alternative scenario for liquidity crises, we show that our framework allows one to infer the shape of latent (unobservable) liquidity from real (observable) order book data. This is important because the concept of a latent order book is sometimes criticized as a figment of the theorist's imagination. Having more direct indications of its existence is comforting.    {For an interesting discussion on the effects of liquidity timescales, see also \cite{corradi2015liquidity}  -- where the authors find that the delayed revelation of latent liquidity breaks the dynamical equilibrium between the market order flow and the limit order flow, triggering  large price jumps}.
\\
 
In section~\ref{sect:pbm_stat} we present the model and derive the governing equations. In section~\ref{sec:stat_states} we compute the stationary states of the limit order book both analytically and numerically. {In section~\ref{sec:calib_stability} we calibrate our model to the order books of over a hundred assets and discuss market stability as function of incentive to reveal and conversion rate}. In section~\ref{sec:impact} we compute the price impact of a metaorder and discuss its behavior.  In section~\ref{sec:ccl} we conclude.

\section{A mechanism for latent liquidity revealing}
\label{sect:pbm_stat}

Our starting point is the reaction-diffusion latent order book model of Donier \emph{et al.} \cite{donierLLOB} (see also \cite{smith2003statistical}). In their setup, the latent volume densities of limit orders in the order book $\rho_\textrm{B}(x,t)$ (bid side) and $\rho_\textrm{A}(x, t)$ (ask side) at price $x$ and time $t$ evolve according to the following rules. Latent orders diffuse with diffusivity constant $D$, are canceled with rate $\nu$, and new intentions are deposited with rate $\lambda$. When a buy intention meets a sell intention they are instantaneously matched: A$\,+\,$B\,=\,$\varnothing$. The trade price $p_\textrm{t}$ is conventionally defined through the equation $\rho_\textrm{B}(p_\textrm{t},t)=\rho_\textrm{A}(p_\textrm{t}, t)$. Donier  \emph{et al.} showed that the resulting stationary order book is locally linear (around the price). In particular, in the infinite memory limit $\nu,\lambda \to 0$ while keeping $\mathcal L = \lambda/\sqrt{\nu D}$ constant, one obtains $\rho_\textrm{A}^\textrm{st}(\xi) =\rho_\textrm{B}^\textrm{st}(-\xi)= \mathcal L \xi   $ for $\xi = x-p_t>0$.\footnote{{Following \cite{donierLLOB} we are setting in the reference frame of the informational price component $\hat{p}_t = \int_0^t ds \, V_s$, where $V_t$ is an exogenous term responsible for the price moves related to common information.}} The parameter $\mathcal L$ was coined the \emph{latent liquidity}. Here we choose to work in such an infinite memory limit. For an analysis of finite memory effects $\nu,\lambda\neq 0$ see \cite{benzaquen2018market}.\\

\begin{figure}[t!]
  \centering
    \includegraphics[width=0.5\columnwidth]{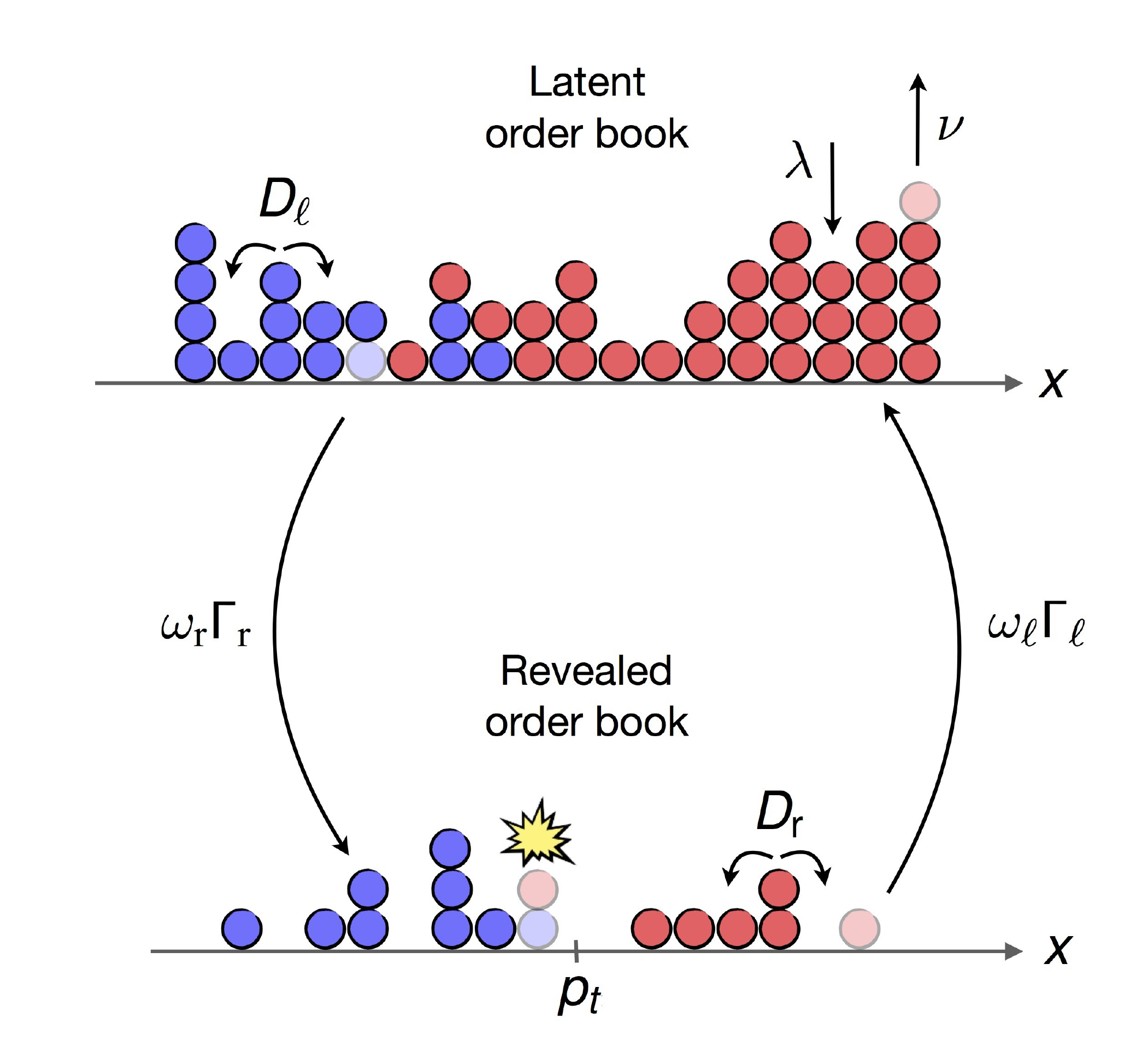}
    \caption{Schematics of the latent and revealed order books. Depositions, cancellations, diffusion and liquidity flow between the two books is signified with arrows (see Sect.~\ref{sect:pbm_stat}).}
    \label{imm:sketch}
\end{figure}

 With the intention of building a mechanism for latent liquidity revealing we define the revealed and latent order books (see Fig.~\ref{imm:sketch}), together with the revealed and latent limit order densities for the bid and ask sides of the book $\rho_\textrm{A/B}^\textrm{(r)}(x,t)$ and $\rho_\textrm{A/B}^{(\ell)}(x,t)$. We denote $D_\ell$ and $D_\textrm{r}$ the diffusion coefficients in the latent and revealed order books respectively. {Here the latent liquidity parameter is naturally defined with respect to the latent diffusivity $D_\ell$, $\mathcal L := \lambda/\sqrt{\nu D_\ell}$ with $\nu,\lambda$ the cancellation and deposition rates in the latent order book (and $\nu,\lambda \to 0$)}. While diffusion in the latent book signifies heterogeneous reassessments of agents intentions \cite{donierLLOB}, the idea of diffusion in the revealed order book certainly deserves a discussion. Once a limit order is placed in the revealed order book, if one wants to change its position before it gets executed, one has to cancel it and place it somewhere else. So one may argue that there should be no diffusion in the revealed book such that revealed order reassessments must go through the latent  book before they can be posted again. However, we believe that $(i)$ unrevealing one's order because one is no longer confident about one's reservation price and waiting for an arbitrary amount of time (of order $(\Gamma \omega_\textrm{r})^{-1}$, see below) to reveal it back, and $(ii)$ canceling an order knowing that it will immediately be posted back at a revised price, are in fact two distinct processes. We thus leave the possibility for a nonzero diffusivity $D_\textrm{r}$ in the revealed order book.
In addition, note that trading fees and priority queues discourage traders from changing posted orders. Indeed, when canceling a revealed order traders lose their position in the priority queue which may result in harmful delays in the execution. One is thus tempted to surmise that $D_\textrm{r} < D_\ell $ in the general case, but one might as well argue that the presence of HFT can considerably increase the value of $D_\textrm{r}$ thereby inverting such an inequality.  Be as it may, the limit $D_\textrm{r}\to 0$ will very likely be an interesting one to address. \\

Furthermore we posit that latent orders are revealed at a position dependent rate $\omega_\textrm{r} \Gamma_\textrm{r}(k \xi)$ and unrevealed at rate $\omega_\ell \Gamma_\ell(k \xi)$, where $\xi = x - p_\textrm{t}$ denotes the distance from the trade price, $k^{-1}$ is a characteristic price scale, and $\Gamma_\textrm{r}$ and $\Gamma_\ell$ are functions taking values in $[0,1]$.\footnote{While it may be reasonable that, when unrevealed, orders don't land on the same price $x$ they took off from (by that moving away the earlier decision, say $(x)_\textrm{r}\to (x+\Delta x)_\ell$, with $\textrm{sign}(\Delta x) = \textrm{sign}(x-p_\textrm{t})$), we shall not consider such a possibility in the present study for reasons of analytical tractability (restrict to $\Delta x = 0$).} Naturally,  buy/sell order matching  $A+B\to \varnothing$ (with rate $\kappa$) only takes place in the revealed order book. Assuming $\Gamma_{\textrm{r}/\ell}$ to be continuous and sufficiently regular on $\mathbb R^*$, one may write for the ask side:
\begin{subeqnarray}
\slabel{eq:ask initial system:1}
\partial_t \rho_\textrm{A}^{(\textrm{r})} &=& D_\textrm{r} \partial_{xx} \rho_\textrm{A}^{(\textrm{r})} + \omega_\textrm{r} \Gamma_\textrm{r} (k \xi) \rho_\textrm{A}^{(\ell)} - \omega_\ell \Gamma_\ell(k \xi) \rho_\textrm{A}^{(\textrm{r})}  - \kappa \rho_\textrm{A}^{(\textrm{r})}\rho_\textrm{B}^{(\textrm{r})}\\ 
\slabel{eq:ask initial system:2}
\partial_t \rho_\textrm{A}^{(\ell)} &=& D_\ell \partial_{xx} \rho_\textrm{A}^{(\ell)} -  \omega_\textrm{r} \Gamma_\textrm{r}(k \xi) \rho_\textrm{A}^{(\ell)} + \omega_\ell \Gamma_\ell(k \xi) \rho_\textrm{A}^{(\textrm{r})}  \ , 
\label{eq:ask initial system}
\end{subeqnarray}
and for the bid side: 
\begin{subeqnarray}
\partial_t \rho_\textrm{B}^{(\textrm{r})} &=& D_\textrm{r} \partial_{xx} \rho_\textrm{B}^{(\textrm{r})} + \omega_\textrm{r} \Gamma_\textrm{r}(-k \xi) \rho_\textrm{B}^{(\ell)} -\omega_\ell \Gamma_\ell(-k \xi) \rho_\textrm{B}^{(\textrm{r})}  -\kappa \rho_\textrm{A}^{(\textrm{r})}\rho_\textrm{B}^{(\textrm{r})}\slabel{eq:bid initial system:1}
\\
\partial_t \rho_\textrm{B}^{(\ell)} &=& D_\ell \partial_{xx} \rho_\textrm{B}^{(\ell)}-   \omega_\textrm{r} \Gamma_\textrm{r}(-k \xi) \rho_\textrm{B}^{(\ell)} + \omega_\ell \Gamma_\ell(-k \xi) \rho_\textrm{B}^{(\textrm{r})}  \ . \slabel{eq:bid initial system:2}
\label{eq:bid initial system}
\end{subeqnarray}
In the limit $\kappa \to \infty$,  $\rho_\textrm{A}^{(\textrm{r})}(x,t)$ and $\rho_\textrm{B}^{(\textrm{r})}(x,t)$ do not overlap  such that one may instead consider the {difference} function $\phi_\textrm{r}(x,t) := \rho_\textrm{B}^{(\textrm{r})}(x,t) - \rho_\textrm{A}^{(\textrm{r})}(x,t)$ and absorb the reaction terms without loss of information. Note however that the bid and ask sides of the latent order book are perfectly allowed to overlap. The trade price $p_\textrm{t}$ is defined as the sign changing point of $\phi_\textrm{r}$:
\begin{equation}
\lim_{\epsilon \to 0}\left[\phi_\textrm{r}(p_\textrm{t}+\epsilon,t) \phi_\textrm{r}(p_\textrm{t}-\epsilon,t) \right] < 0 \ .
\label{eq:trade price definition}
\end{equation}
For the sake of simplicity we shall set $\omega_\textrm{r} = \omega_\ell = \omega$.\footnote{Note that relaxing this hypothesis is possible ({see e.g. Sec.~\ref{analnumsols} for the case $D_\textrm{r}=0$}) but at the cost of additional complexity. We checked that this point was without effect on our main qualitative results.\smallskip} We also set $\Gamma_\textrm{r} = 1 - \Gamma_\ell := \Gamma$ where $\Gamma$ is the conversion probability distribution function for which we impose $\Gamma(y \leq 0)=1$. This implies that latent orders falling on the "wrong side" reveal themselves with rate $\omega$. One may rightfully argue that such orders should then be executed against the best quote, consistent with real market rules for which the real order book cannot be crossed. However this would prevent analytical progress. We ran numerical simulations (see section \ref{numsim}) in which such revealed orders are properly located at the best quote and did not observe any significant impact on our main results. \\

Subtracting Eq.~\eqref{eq:ask initial system:1} to Eq.~\eqref{eq:bid initial system:1} and injecting 
$\rho_\textrm{B}^{(\textrm{r})} = \phi_\textrm{r} \mathds{1}_{\{ x < p_\textrm{t} \}}$, $\rho_\textrm{A}^{(\textrm{r})} = -\phi_\textrm{r} \mathds{1}_{\{ x > p_\textrm{t} \}}$, one obtains the following set of equations, central to our study: 
\begin{subeqnarray}
 \displaystyle \partial_t \rho_\textrm{B}^{(\ell)} &=& D_\ell \partial_{xx} \rho_\textrm{B}^{(\ell)} -  \omega \left\{ \Gamma(-k \xi)\rho_\textrm{B}^{(\ell)} - \left[1 - \Gamma(-k \xi) \right] \mathds{1}_{\{ x < p_\textrm{t} \}} \phi_\textrm{r}  \right\} \slabel{lorenzo1}
\\
 \displaystyle \partial_t \rho_\textrm{A}^{(\ell)}  &=& D_\ell \partial_{xx} \rho_\textrm{A}^{(\ell)} -  \omega \left\{ \Gamma(k \xi)\rho_\textrm{A}^{(\ell)} +  \left[1 - \Gamma(k \xi) \right] \mathds{1}_{\{ x > p_\textrm{t} \}} \phi_\textrm{r}  \right\} \slabel{lorenzo2}
\\
 \displaystyle \partial_t \phi_\textrm{r} &=& D_\textrm{r} \partial_{xx} \phi_\textrm{r} +    \omega \left\{\Gamma(-k \xi) \rho_\textrm{B}^{(\ell)} - \Gamma(k \xi) \rho_\textrm{A}^{(\ell)} -\left[1 - \Gamma(k |\xi|) \right] \phi_\textrm{r}  \right\}\ . 
\label{eq:rhophi1}\slabel{lorenzo3}
\end{subeqnarray}
Equations~\eqref{eq:rhophi1}  must be complemented with a set of boundary conditions. 
 In particular we impose that $\lim_{x\to \infty} \partial_x\rho^{{(\ell)}}_\textrm{A} = - \lim_{x\to - \infty} \partial_x\rho^{{(\ell)}}_\textrm{B} = \mathcal L$ (see \cite{donierLLOB}), and that $ \rho^{{(\ell)}}_\textrm{A}$ when $x\to - \infty$, respectively $  \rho^{{(\ell)}}_\textrm{B}$ when $x\to \infty$, do not diverge. In addition, whenever $D_\textrm{r} \neq 0$,\footnote{While $D_\textrm{r} = 0$ is an interesting limit,  $D_\ell = 0$ does not seem to be particularly appealing on modelling grounds. In section \ref{analnumsols} we give a more solid theoretical argument to support this claim.
 } one must impose that $\displaystyle \phi_\textrm{r}(0) = 0$ and $\phi_\textrm{r}(x) $ does not diverge  when $|x|\to  \infty$.

\section{Stationary order books}
\label{sec:stat_states}

In this section we compute analytically and numerically the stationary order books as function of the different parameters, and discuss interesting limit cases. Setting $\partial_t \rho_\textrm{B}^{(\ell)} = \partial_t \rho_\textrm{A}^{(\ell)}=\partial_t \phi_\textrm{r} =0$ in Eqs.~\eqref{eq:rhophi1} one  obtains for all $  \xi \in \mathbb{R}^*, \  \rho_\textrm{B}^{(\ell)} (\xi) = \rho_\textrm{A}^{(\ell)} (-\xi)$ and $ \phi_\textrm{r} (\xi) = - \phi_\textrm{r} (-\xi)$. This allows one to solve the problem on \( \mathbb{R}^{+*} \), with boundary conditions $\displaystyle  \rho_\textrm{B}^{(\ell)} (0^+) = \rho_\textrm{A}^{(\ell)} (0^+), \partial_{\xi}\rho_\textrm{B}^{(\ell)} (0^+) = - \partial_\xi\rho_\textrm{A}^{(\ell)}(0^+)$. The system one must solve for $\xi>0$ reduces to:
 \begin{subeqnarray}
\slabel{eq:general system for choice of gamma:1}
 0 &=& D_\ell\partial_{\xi\xi} \rho_\textrm{B}^{(\ell)} - \omega \rho_\textrm{B}^{(\ell)}  \\
\slabel{eq:general system for choice of gamma:2}
 0 &=&D_\ell\partial_{\xi\xi} \rho_\textrm{A}^{(\ell)} - \omega \left\{\Gamma(k\xi)\rho_\textrm{A}^{(\ell)}+[1-\Gamma(k\xi)]\phi_\textrm{r}\right\}\\
\slabel{eq:general system for choice of gamma:3}
 0 &=&D_\textrm{r}\partial_{\xi\xi} \phi_\textrm{r} - \omega \left\{\Gamma(k\xi)\rho_\textrm{A}^{(\ell)}+[1-\Gamma(k\xi)]\phi_\textrm{r} - \rho_\textrm{B}^{(\ell)}\right\}  \ .
\label{eq:general system for choice of gamma}
\end{subeqnarray}

\subsection{Analytical and numerical solutions}
\label{analnumsols}

Here we provide a solution of Eqs.~\eqref{eq:general system for choice of gamma} for three distinct cases of interest $D_\textrm{r} = 0 $, $D_\textrm{r} = D_\ell $, and $D_\textrm{r} \neq D_\ell$. Note that for  $D_\ell = 0$, one can show that  $\rho_\textrm{B}^{(\ell)}(\xi > 0) = 0$, while $\phi_\textrm{r}(\xi > 0) = a\xi + b$ with $a$ and $ b$ two constants, and $\rho_\textrm{A}^{(\ell)} = \phi_\textrm{r}[\Gamma(k\xi)-1]/\Gamma(k\xi) \approx -\phi_\textrm{r}/\Gamma(k\xi)$ for $\xi \to \infty$, by assuming $\Gamma$ to vanish at infinity. Such solutions are not compatible with the boundary conditions at infinity and will therefore not be further inspected.

\subsubsection*{The limit case \boldmath$D_\textrm{r} = 0$}

Setting $D_\textrm{r} = 0 $ in Eqs.~\eqref{eq:general system for choice of gamma} and introducing  $ \ell_\ell := \sqrt{{D_\ell}/{\omega}}$, on finds that the stationary order book densities are given by (see Fig.~\ref{imm:stat_books}(a)):
\begin{subeqnarray}
\displaystyle  \rho_\textrm{B}^{(\ell)} (\xi) &=& \frac{\mathcal{L}\ell_\ell}{2} e^{-{\xi}/{\ell_\ell}} \slabel{rhobb} \\
\displaystyle \rho_\textrm{A}^{(\ell)} (\xi) &=& \mathcal{L} \xi + \frac{\mathcal{L}\ell_\ell}{2} e^{{-\xi}/{\ell_\ell}} \slabel{rhoaa} \\
\displaystyle \phi_\textrm{r} (\xi) &=& \frac{\mathcal{L}\ell_\ell}{2} e^{-{\xi}/{\ell_\ell}} - \frac{\mathcal{L} \xi \Gamma (k\xi) }{1 - \Gamma (k\xi)}  \ .\slabel{phirr}
\label{eq: sol for Dr=0}
\end{subeqnarray}
Let us stress that the solution $\phi_\textrm{r} (\xi)$ is discontinuous in $\xi=0$, consistent with no diffusion in the revealed order book. More precisely, provided $\Gamma'(0^+) \neq 0$, one has $\phi_\textrm{r} (0^-) - \phi_\textrm{r} (0^+)= - \mathcal{L} \left[\ell_\ell+{2}/({\Gamma'(0^+)k} )\right]$.
While the solution for the case $D_\textrm{r} = 0$ can be expressed for an arbitrary function $\Gamma(y>0)$, this is not the case for $D_\textrm{r} \neq 0$ and one must specify its shape. In the following we choose to work with:
\begin{equation}
\Gamma (y) \ =\
\Bigg\{
\renewcommand*{\arraystretch}{2}
\begin{array}{clr}
 1 &&\forall y \leq 0 \vspace{-0.2cm}\\
e^{-y} &&\forall y > 0 \ .
\end{array}
\label{eq:phi limit}
\end{equation}
Note that studying the effect of a scale-invariant power law decaying $\Gamma$ could also yield interesting results.\footnote{Also note that in this particular case, the solution for $\omega_\ell\neq \omega_\textrm{r}$ can be easily expressed by substituting $\ell_\ell$ by $\ell_\ell^{(\textrm{r})} = \sqrt{{D_\ell}/{\omega_{\textrm{r}}}}$ into Eqs.~\eqref{rhobb} and \eqref{rhoaa}, and replacing Eq.~\eqref{phirr}
by:
$$
	\phi_\textrm{r}(\xi) = \frac{\omega_\textrm{r}}{\omega_\ell}\left[\frac{\mathcal{L}\ell_\ell^{(\textrm{r})}}{2}e^{-\xi/\ell_\ell^{(\textrm{r})}} - \frac{\mathcal{L}\xi e^{-k\xi}}{1-e^{-k\xi}} \right] \ .
$$}

\subsubsection*{The limit case \boldmath $D_\textrm{r} = D_\ell$}
\begin{figure}[t]
        \includegraphics[width=\textwidth]{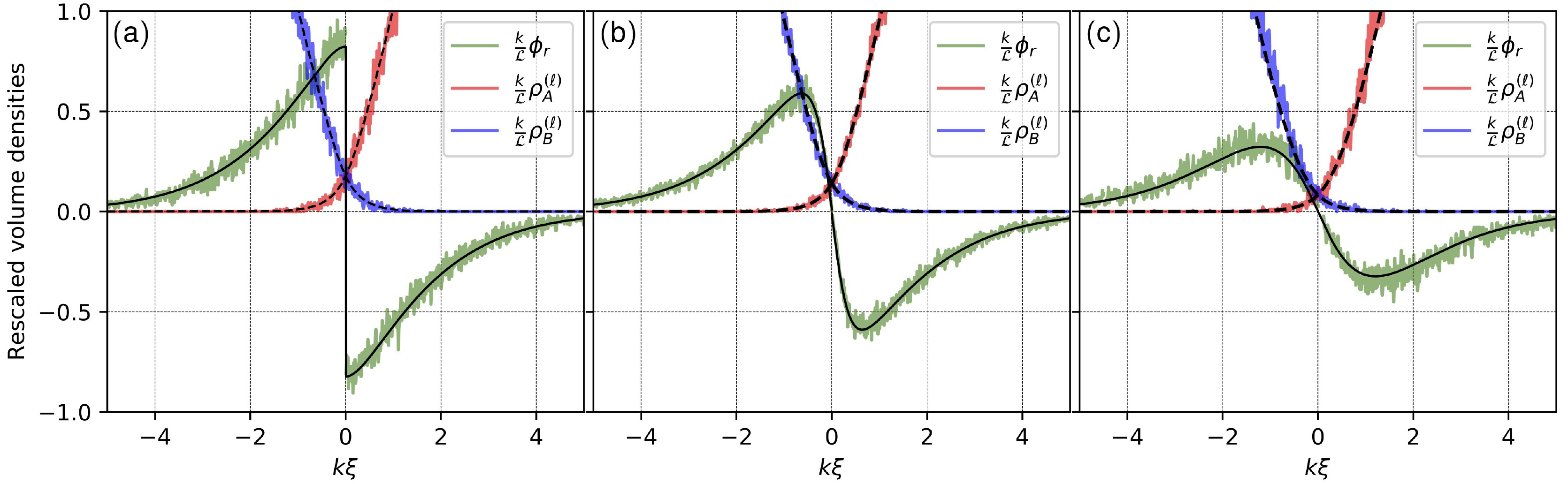}
        \caption{ Rescaled stationary order books as function of rescaled price for (a) $D_\textrm{r} = 0 $, (b) $D_\textrm{r} \neq D_\ell $ ($D_\textrm{r} < D_\ell $), and (c) $D_\textrm{r} = D_\ell$. Solid black lines indicate the rescaled {theoretical} revealed order density $ k \phi_\textrm{r}/\mathcal L$ while dashed black lines signify the {theoretical} latent order densities $k\rho_\textrm{A/B}^{(\ell)}/\mathcal L$. The results of the numerical simulation presented in Sect.~\ref{numsim} are plotted with color lines on top of the analytical curves, with  $k\ell_\ell = 0.35$ for (a) and (c), and $ \ell_\textrm{r}/\ell_\ell = 0.32$ for (b). {Typical computational time: 1\,min.}}
          \label{imm:stat_books}
\end{figure}

For $D_\textrm{r} = D_\ell$ and provided $k\ell_\ell \neq 1$  the stationary books are given by (see Fig. \ref{imm:stat_books}(c)):
\begin{subeqnarray}
	\slabel{eq:solution for Dr=Du:1}
	\rho_\textrm{B}^{(\ell)} (\xi) &=& \frac{\mathcal{L}}{k} g(k\ell_\ell) e^{- {\xi}/{\ell_\ell} } \slabel{rhobbb}
	\\
	\slabel{eq:solution for Dr=Du:2}
	\rho_\textrm{A}^{(\ell)} (\xi) &=& \mathcal{L} \xi + \frac{\mathcal{L}}{k} g(k\ell_\ell) e^{- {\xi}/{\ell_\ell} } +  \phi_\textrm{r} (\xi)
	\\
	\slabel{eq:solution for Dr=Du:3}
	\phi_\textrm{r} (\xi) &=& \mathcal{L} \left( \frac{1}{(k\ell_\ell)^2-1} \left[\xi + \frac{2k\ell_\ell^2}{(k\ell_\ell)^2-1} \right] e^{-k\xi} + \frac{g(k \ell_\ell)}{k^2 \ell_\ell(k \ell_\ell + 2)} e^{-(1/ \ell_\ell+k)\xi}  \right.\nonumber\\ 
	&&  +  \left.\left[\frac{g(k\ell_\ell)}{2k\ell_\ell}\xi   - \frac{2k \ell_\ell^2}{[(k \ell_\ell)^2-1]^2}- \frac{g(k \ell_\ell)}{k^2\ell_\ell(k\ell_\ell+2)} \right]e^{-{\xi}/{\ell_\ell}} \right) \ ,   \slabel{phirrr}
	\label{eq:solution for Dr=Du}
\end{subeqnarray}
where we introduced ${g}(\zeta) = {2\zeta^2(2+\zeta)^2}/{[(1+\zeta)^2(8+3\zeta)]}$. For $k\ell_\ell = 1$ the result can be obtained by Taylor expanding Eqs.~\eqref{eq:solution for Dr=Du}  about $k\ell_\ell = 1$. Note that in this case the function $\phi_\textrm{r}(\xi)$ is continuous in $\xi=0$, consistent with nonzero diffusivity in the revealed order book.\\

\subsubsection*{The general case \boldmath$D_\textrm{r} \neq D_\ell$}

For $D_\textrm{r} \neq D_\ell$ the set of Eqs.~\eqref{eq:general system for choice of gamma} must be solved numerically. Figure~\ref{imm:stat_books}(b) displays a plot of the stationary order books computed using a finite difference method for $D_\textrm{r}/D_\ell\approx 0.1$, that is $\ell_\ell/\ell_\textrm{r}\approx 3$ where we introduced $\ell_\textrm{r} := \sqrt{{D_\textrm{r}}/{\omega}}$.  As one can see, in the situation $D_\textrm{r} < D_\ell$ the revealed order book's shape is somewhat in between the cases $D_\textrm{r} =0$ and $D_\textrm{r} = D_\ell$, that is continuous but with a steeper slope at $\xi = 0$. As expected, a little amount of diffusion in the revealed order book suffices to  regularize the singularity at the trade price.\\

 As it can be seen from Eq.~\eqref{eq:general system for choice of gamma:1} (or from Eqs.~\eqref{rhobb} and \eqref{rhobbb} in the particular cases $D_\textrm{r} = 0$ and $D_\textrm{r} = D_\ell$), in all cases $\ell_\ell$ denotes the typical scale over which the latent books overlap. This is consistent with the idea that $\ell_\ell$ is the typical displacement by diffusion of a latent order in the vicinity of the trade price during a time interval $\omega^{-1}$, that is before it gets revealed. Also, it can be seen from Eqs.~\eqref{phirr} and \eqref{phirrr} in the particular cases $D_\textrm{r} = 0$ and $D_\textrm{r} = D_\ell$ that the typical horizontal extension of the revealed order book, commonly called \emph{order book depth}, is given by $\max(k^{-1},\ell_\ell)$, consistent with the decay 
of the conversion probability function $\Gamma$ and the horizontal extension of the latent books. Note however that, as shall be argued in Sect.~\ref{sec:calib_stability}, $k^{-1}$ must always be of order or larger than $\ell_\ell$ for stability reasons, and therefore $\max(k^{-1},\ell_\ell) \sim k^{-1}$. In the following we shall thus call $k^{-1}$ the order book depth. Finally note that when $\ell_\textrm{r}$ (equivalently $D_\textrm{r}$) is decreased while keeping all other parameters constant, the slope 
of the revealed order book around the origin increases, by that concentrating further the available liquidity around the trade price. \\

\subsection{The LLOB limit}

Our model being built upon the locally linear order book model (LLOB) by Donier \textit{et al.}  \cite{donierLLOB}, we should be able to recover such a limit for certain values of the parameters. Since there is only one diffusion coefficient in the LLOB model, the latter should correspond to the  $D_\textrm{r} = D_\ell$ case. Then, the LLOB model assumes no lag effect, i.e. latent orders are immediately executed when at the trade price. This translates into $\omega\to \infty$ or equivalently $\ell_\ell\to 0$, that is no overlap of the latent books. 
More rigorously, nondimensionalizing Eqs.~\eqref{eq:solution for Dr=Du} as $\{\tilde{\phi_\textrm{r}}, {\tilde\rho^{(\ell)}}\}= \{{k \phi_\textrm{r}}/{\mathcal{L}}, {k \rho^{(\ell)}}/{\mathcal{L}}\}$ and $\tilde{\xi} = k\xi$, one can see that taking the limit  $k \ell_\ell\to 0$ yields for all $\tilde \xi >0$, $\tilde \rho_\textrm{B}^{(\ell)}(\tilde \xi) \to 0$, and 
 $\tilde \rho_\textrm{A}^{(\ell)}(\tilde \xi) + \tilde  \rho_\textrm{A}^{(\textrm{r})}(\tilde \xi) = \tilde \xi := \tilde \rho_\textrm{A}^\textrm{LLOB}(\tilde  \xi)$ that is precisely the LLOB result. To summarize, provided the latent order book of Donier \textit{et~al.} is defined as the sum of the latent and revealed books, the LLOB limit is recovered for $k \ell_\ell \ll 1$. The latter condition indicates that the typical displacement $\ell_\ell$ of latent orders in the vicinity of the price must remain small compared to the order book depth $k^{-1}$. Note that, despite what a first intuition might suggest, the condition $k\to \infty$, that is no incentive to give away information until absolutely necessary, is not required to recover the LLOB limit.

\subsection{Numerical simulation}
\label{numsim}

In order to test our results, we performed a numerical simulation of our model (see Fig.~\ref{imm:stat_books}) which proceeds as follows. We define four vectors $\rho_\textrm{A}^{(\ell)},\rho_\textrm{A}^{(\textrm{r})},\rho_\textrm{B}^{(\ell)},\rho_\textrm{B}^{(\textrm{r})}$ of size 2000 on the price axis (a good trade-off between the continuous approximation of the model and computational time cost). Each component of vectors stores the number of orders contained at that price at a given time. At each cycle we draw the orders that shall diffuse from a binomial distribution of parameter $p_{\ell/\textrm{r}}$, directly related to the diffusion constant in the respective book by the relation $D_{\ell/\textrm{r}} = p_{\ell/\textrm{r}}/(2 \tau)$ where $\tau$ denotes the time step. Some of the orders (drawn from a binomial of parameter ${1}/{2}$) will move to the left and the remaining to the right. Reflecting boundary conditions are imposed for revealed orders; the slope of the latent book at the boundaries is ensured by an incoming current of particles $J = D_\ell \mathcal{L}$. Then, some orders in the latent book are drawn from a binomial distribution of parameter $\omega \tau \Gamma(k \xi)$ for the ask side (resp. ${\omega \tau}\Gamma(-k \xi)$ for the bid side)) and moved to the revealed book. Here $p_\textrm{t}$ denotes the mid-price. Equivalently revealed orders are moved to the latent book, only with parameter $\omega \tau  (1-\Gamma(\cdot))$.
 Whenever bid and ask orders are found at the same price, they are cleared from the book. Figure~\ref{imm:stat_books} shows that the results of the numerical simulations are in very good agreement with the analytical solutions.

\section{Market stability and calibration to real data}
\label{sec:calib_stability}

In this section, we address the important question of market stability, as given by the amount of liquidity in the revealed order book. Clearly, when the conversion rate $\omega$ is  low, the revealed liquidity is thin and prices can be prone to liquidity crises, even when the latent liquidity is large. We calibrate our model to real order book data and discuss the results in the light of the stability map provided by our model.

\subsection{Market stability}

Imposing that the order densities $\rho_\textrm{A}^{(\ell)},\rho_\textrm{A}^{(\textrm{r})},\rho_\textrm{B}^{(\ell)},\rho_\textrm{B}^{(\textrm{r})}$ must be non negative, consistent with a physically meaningful solution, restricts the possible values of $\ell_\ell := \sqrt{{D_\ell}/{\omega}}$ for a given order book depth $k^{-1}$. \\
 
In the  $D_\textrm{r} = 0 $ case, combining Eq.~\eqref{phirr} with  Eq.~\eqref{eq:phi limit}  yields $\phi_\textrm{r}(0^+)  = \mathcal{L} \left[ \ell_\ell/2 - {1}/{k} \right]$. Restricting to $\phi_\textrm{r}(0^+)\leq 0$ (which is tantamount to $\rho_\textrm{A}^{(\textrm{r})} (0^+),\rho_\textrm{B}^{(\textrm{r})}(0^+) \geq 0 $)  gives $k\ell_\ell \leq 2 $. Note however that  this condition is not sufficient to say that the order densities are everywhere positive, but that they are only positive around the origin. The necessary and sufficient condition to ensure full positiveness reads $k \ell_\ell\leq 1$. For $1 \leq k \ell_\ell \leq 2$ the order book displays a "hole" along the price axis, but is well defined around the origin. 
  Since we are most interested in the revealed liquidity in the vicinity of the trade price we will choose $k \ell_\ell \leq 2$ (or, in terms of $\omega$, $\omega \geq D_\ell k^2/4$) as our \emph{stability condition}, with no qualitative and only little quantitative effect on our main conclusions.  
The maximum amplitude of the real order book density scales as:
\begin{equation}
\max_{\xi } \big | \phi_\textrm{r}(\xi)\big | = -\phi_\textrm{r}(0^+)= \frac{\mathcal{L}}{k} \left(1-\frac{k\ell_\ell}{\zeta_\textrm{c}} \right), \quad \zeta_\textrm{c} = 2 \ .
\label{eq: scaling Dr0}
\end{equation}

For $D_\ell = D_\textrm{r}$ the stability condition is imposed by the sign of the slope at $\xi=0$. One finds that the critical value of $k\ell_\ell$ for which the liquidity around the origin vanishes is given by $\zeta_\textrm{c} =  [-2 + (73-6\sqrt{87} )^{{1}/{3}} + (73+6\sqrt{87} )^{{1}/{3}} ]/3\approx 1.875$.\footnote{Note that in this case the solution, provided $k \ell_\ell > 1$, is also asymptotically unstable: $ \lim_{\xi \to \infty} \phi_\textrm{r}(\xi) = \mathcal{L} {g(k\ell_\ell)}\xi e^{-{\xi}/{\ell_\ell}}/({2k\ell_\ell}) > 0$, {while  for $k\ell_\ell = 1$, $\lim_{\xi \to \infty} \phi_\textrm{r} < 0 $}. 
}
 Arguing that in this case the maximum of the density can be approximated by $\max_{\xi}\big |\phi_\textrm{r}'(0^+)\xi e^{-k\xi}\big |= |\phi_\textrm{r}'(0^+)|/(ek)$ yields: 
\begin{equation}
\max_{\xi } \big | \phi_\textrm{r}(\xi)\big |\sim  \frac{\mathcal{L}}{ek} \frac{\zeta_\textrm{c}(3\zeta_\textrm{c}^2+4\zeta_\textrm{c}-3)}{(1+\zeta_\textrm{c})^2(8+3\zeta_\textrm{c})} \left(1-\frac{k\ell_\ell}{\zeta_\textrm{c}} \right) , \quad \zeta_\textrm{c} \approx 1.875 \ . 
\label{eq:scaling DrDu}
\end{equation}
\begin{figure*}[t!]
        \centering
        \includegraphics[width=0.99\columnwidth]{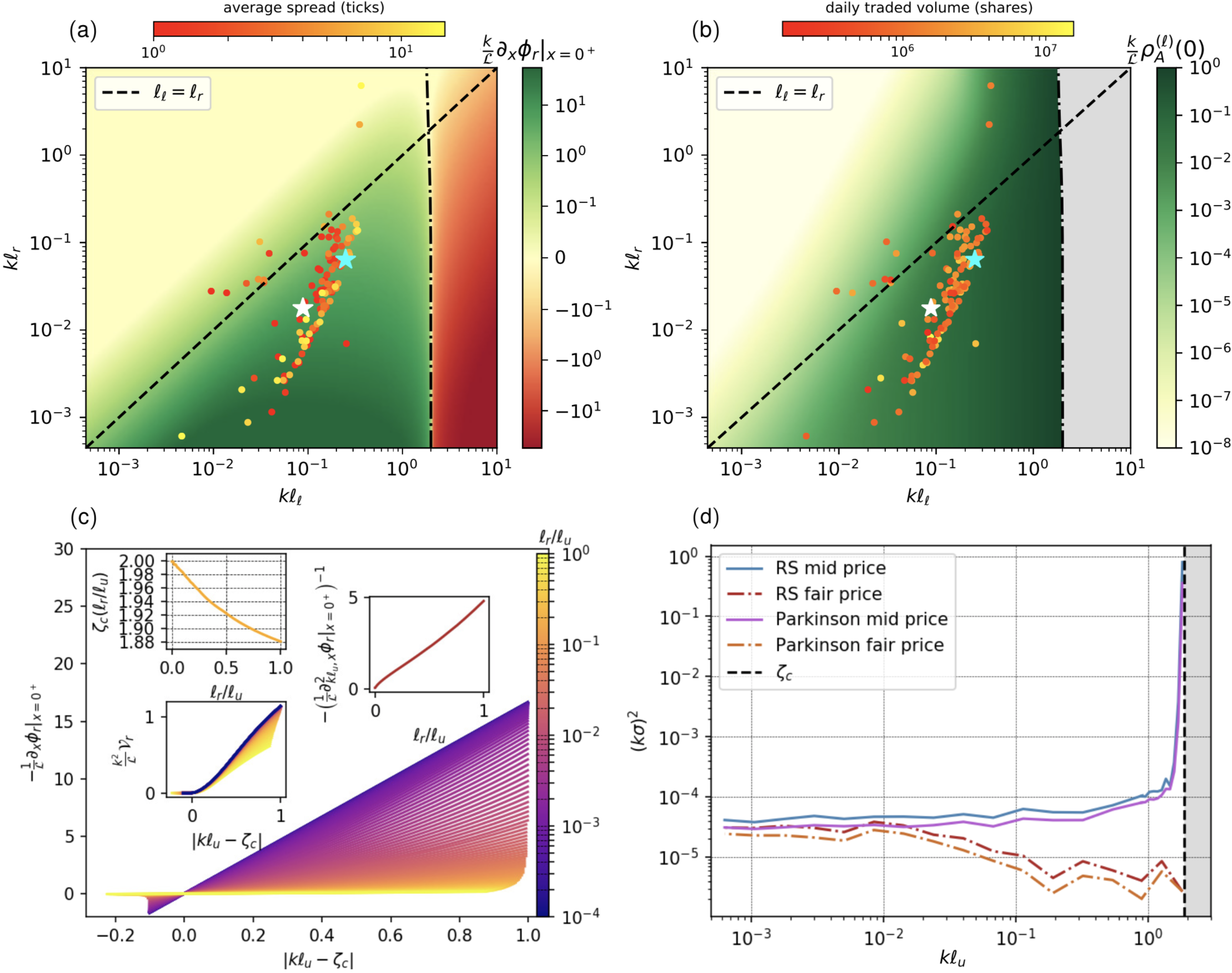}
        \caption{Parametric study of stationary order books. (a) Stability map. Density plot (symlog scale) of the rescaled slope of the  revealed order density at the origin, as function of $k \ell_\ell$ and $k \ell_\textrm{r}$. The dashed line indicates $\ell_\textrm{r}= \ell_\ell$, the dash-dotted line indicates the critical line $\phi_\textrm{r}'(0)=0$.         
        (b) Overlap of the latent order books. Density plot (log scale) of the $y$-intercept of the  latent order book densities at the origin. Note that such a quantity is a direct measure of the overlap. {Typical computational time to draw the stability map: 1\,h.} The empirical order book data (US stocks, see Tab.~\ref{tablestocks}, and Euro Stoxx, white star) are reported with a colormap scaling with their average spread value (see (a)) and with their average daily traded volume (see (b)). {The light blue star results from the calibration of Euro Stoxx data 2h around the flash crash of February 5th 2018 at 8:15pm (Paris time)}. {Typical computational time: 1\,min for each asset.}
         (c) Plot of the slope of the revealed order book at the origin as function of  {$ \zeta_\textrm{c}- k \ell_\ell$}, for different values of $\ell_\textrm{r}/\ell_\ell$. The top inset shows $\zeta_\textrm{c}$ as function of $\ell_\textrm{r}/\ell_\ell$, and the center left inset displays the total volume in the revealed order book $\mathcal{V}_\textrm{r} = \big |\int_0^\infty \text d\xi \, \phi_\textrm{r}(\xi)\mathds{1}_{\phi_\textrm{r}(\xi)<0} \big|$,  as function of {$ \zeta_\textrm{c} - k \ell_\ell $}. {The right inset shows the angular coefficient of the slope of the main plot at the origin}.
         (d) Numerical volatility. Plot of rescaled numerical squared volatility of the trade price and the fair price (see Eq.~\eqref{eq: fair price})   in the $\ell_\textrm{r}= \ell_\ell$ case. We display two different estimations: {Rogers-Satchell, $\sigma_{\textrm{RS}}^2 = \mathbb{E}\left[(p_\textrm{H}-p_\textrm{O})(p_\textrm{H}-p_\textrm{C})+(p_\textrm{L}-p_\textrm{O})(p_\textrm{L}-p_\textrm{C})\right]$, and Parkinson, $\sigma_\textrm{p}^2 = \frac{1}{4\ln(2)} \mathbb{E}\left[(p_\textrm{H}-p_\textrm{L})^2\right]$ where $p_\textrm{H}, p_\textrm{L}, p_\textrm{O}, p_\textrm{C}$  denote  the high, low, open and close prices respectively \cite{rogers1994estimating,parkinson1980extreme}.  
    }}
\label{imm:stab}
\end{figure*}

Figure~\ref{imm:stab}(a) displays $-k\phi_\textrm{r}'(0^+)/\mathcal{L}$ as function of the dimensionless parameters $k\ell_\ell$ and $k\ell_\textrm{r}$. The dash-dotted line  corresponding to $\phi_\textrm{r}'(0)=0$ splits the parameter space into a stable region (green) and an unstable region (red). 
We naturally checked that the analytical values of $\zeta_\textrm{c}$ obtained above for $\ell_\ell=\ell_\textrm{r}$ and $\ell_\textrm{r}\to 0$ are recovered.  As one can see, while the role played by $\ell_\textrm{r}$ with respect to the position of the critical line $\zeta_\textrm{c}$ is quite marginal (quasi-vertical dash dotted line),  the slope of the order book around $\xi=0$ increases with decreasing $\ell_\textrm{r}/\ell_\ell $. 
More precisely, Fig.~\ref{imm:stab}(c) displays the slope of the revealed book in the vicinity of the transition and shows that, at given $\ell_\textrm{r}/\ell_\ell$, the slope indeed scales linearly with $| k \ell_\ell - \zeta_\textrm{c}|$.  In addition the top right inset shows that the slope also scales as $\ell_\ell/\ell_\textrm{r}$, which   finally leads to $|\phi_\textrm{r}'(0^+)|\sim | k \ell_\ell - \zeta_\textrm{c}|(\ell_\ell/\ell_\textrm{r})$. The other insets show $\zeta_\textrm{c}$ as function of $\ell_\textrm{r}/\ell_\ell$,  and the total revealed volume.
For the sake of completeness, Fig.~\ref{imm:stab}(b) displays a proxy of the overlap between the latent bid and ask books in the parameter space $(k\ell_\ell,k\ell_\textrm{r})$. While vanishing in the region $k\ell_\ell\ll  \zeta_\textrm{c}$, the overlap is quite large (of the order of $k^{-1}$) in the vicinity of the critical line $k\ell_\ell \lesssim \zeta_\textrm{c}$  indicating a large volume of latent orders in the vicinity of the price. Interestingly, combined with a vanishing level of liquidity, the increased level of activity around the origin  induces important fluctuations of the trade price. To illustrate this,  Fig.~\ref{imm:stab}(d) displays the numerically determined squared volatility of the trade price $p_{t}$ as function of $k\ell_\ell$ in the $\ell_\textrm{r} = \ell_\ell$ limit. For comparison, we also plotted the  volatility of the \emph{fair price} $p^\textrm{f}_t$, here defined as the value that equilibrates total (revealed and latent) supply and demand:
\begin{equation}
\int_{0}^{p^\textrm{f}_t} \textrm d\xi \left[{\rho}_\textrm{A}^{(\textrm{r})}(\xi,t)+{\rho}_\textrm{A}^{(\ell)}(\xi,t)\right] = \int_{p^\textrm{f}_t}^{\infty} \textrm d\xi \left[{\rho}_\textrm{B}^{(\textrm{r})}(\xi,t)+{\rho}_\textrm{B}^{(\ell)}(\xi,t)\right] \ .
\label{eq: fair price}
\end{equation}
As one can see, for $\omega \gg \omega_\textrm{c} := D_\ell k^2 /\zeta_\textrm{c}^2$ the volatility of the trade price coincides with its fair price counterpart, consistent with the idea that for high conversion rates the coupling between the revealed and latent books is almost instantaneous and therefore the mid-price tends to follow the fair price. In the vicinity of the critical line the volatility of the fair price slightly decreases. However, the volatility of the trade price strongly diverges as the vanishing liquidity limit is approached, i.e. when the conversion rate $\omega$ decreases to $\omega_\textrm{c}$. Note that 
while the trade {price} can no longer be defined when a liquidity crisis arises, the fair price as defined in Eq.~\eqref{eq: fair price} remains well behaved.
We also investigated the volatility in the $\ell_\textrm{r} = 0$ limit and obtained similar qualitative results, only with weaker overall volatility levels consistent with liquidity concentration around the trade price. 
\\

At this point, let us summarise our results. The market is most stable when the conversion rate is large and/or the latent orders diffusivity is small. In these cases, there is a good level of revealed liquidity and the trade price follows the fair price. However, when lag effects become important, and more particularly when the conversion time $\omega^{-1}$ becomes longer than the typical diffusion time of a latent order across the revealed order book depth $k^{-1}$, the order book empties out and the trade price undergoes large excursions away from the fair price, {see also \cite{corradi2015liquidity,cristelli2010liquidity}}. As for the effect of diffusion in the real order book, a small $\ell_\textrm{r}$ concentrates the liquidity around the origin, therefore providing a wall to price fluctuations, while a large $\ell_\textrm{r}$ induces a weaker revealed order book slope around the origin, facilitating larger price excursions. Thus, our framework in the vicinity of the critical line could provide an interesting scenario to understand the nature of liquidity crises in financial markets. One indeed expects that in the presence of increased uncertainty, the conversion rate will decrease (as investors remain on the sidelines) and latent orders diffusivity will increase, both effects driving the market towards a liquidity crisis.

\subsection{Order book data}
\label{empiricalsec}

\begin{figure*}[t!]
        \centering
        \includegraphics[width=0.55\columnwidth]{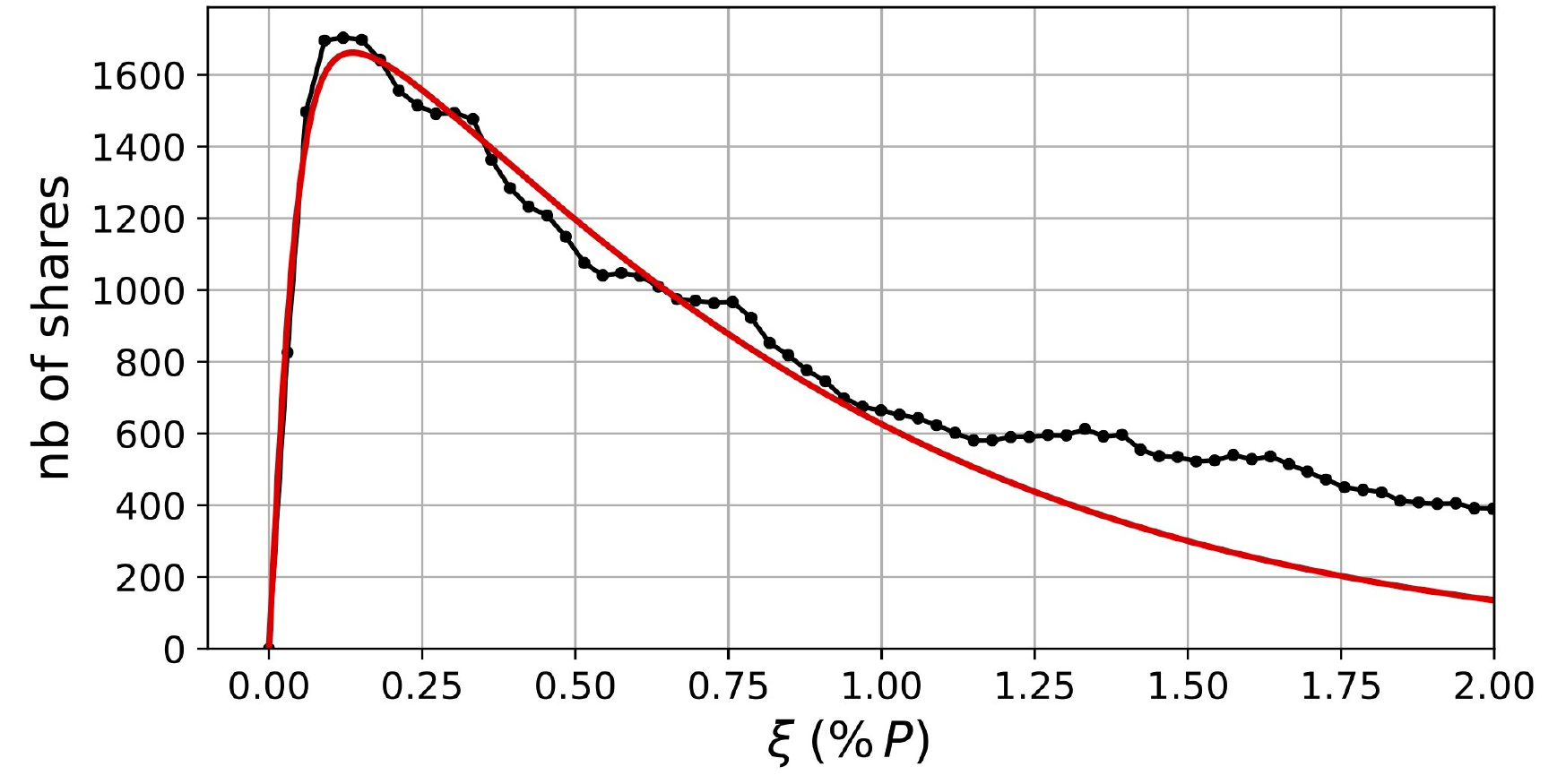}
        \caption{Fit of the stationary revealed order book $\phi_\textrm{r}$ to the average Euro Stoxx futures contract order book. The calibration yields $\mathcal L = 4599$, $k=2.12$, $\ell_\ell = 0.042$ and $\ell_r = 0.0084$. {Typical computational time: 1\,min to obtain the average order book of the asset and apply the fitting procedure.} }
\label{imm:fit}
\end{figure*}

To illustrate how our model allows one to calibrate latent parameters using visible order book data, we calibrate it to the real order book of the Euro Stoxx futures contract and of a set of over 100 large cap US stocks extracted from their primary market on the period August 2017 to  April 2018, see Table~\ref{tablestocks}. For each asset, we take a snapshot of the order book every ten minutes (every minute for the Euro Stoxx contract), on regular trading hours. We then fold the bid side onto the ask side, the origin being taken at the opposite best \cite{bouchaud2018trades}, we average over the whole sample and rescale the $x$-axis by the average price taken over the sample period. Finally, fitting the stationary order book $\phi_\textrm{r}$ (see Figs.~\ref{imm:fit}, \ref{imm:fit_stoks1}, \ref{imm:fit_stoks2} and \ref{imm:fit_stoks3}) outputs four parameters: $\mathcal{L}$, $k$, $\ell_\ell$ and $\ell_\textrm{r}$ (see Table~\ref{tablestocks}). Note that $k^{-1}$, $\ell_\ell$ and $\ell_\textrm{r}$ have units of $\% $p, and $\mathcal{L}$ is in shares per $\% $p.
\\

\begin{figure*}[t!]
        \centering
        \includegraphics[width=0.95\columnwidth]{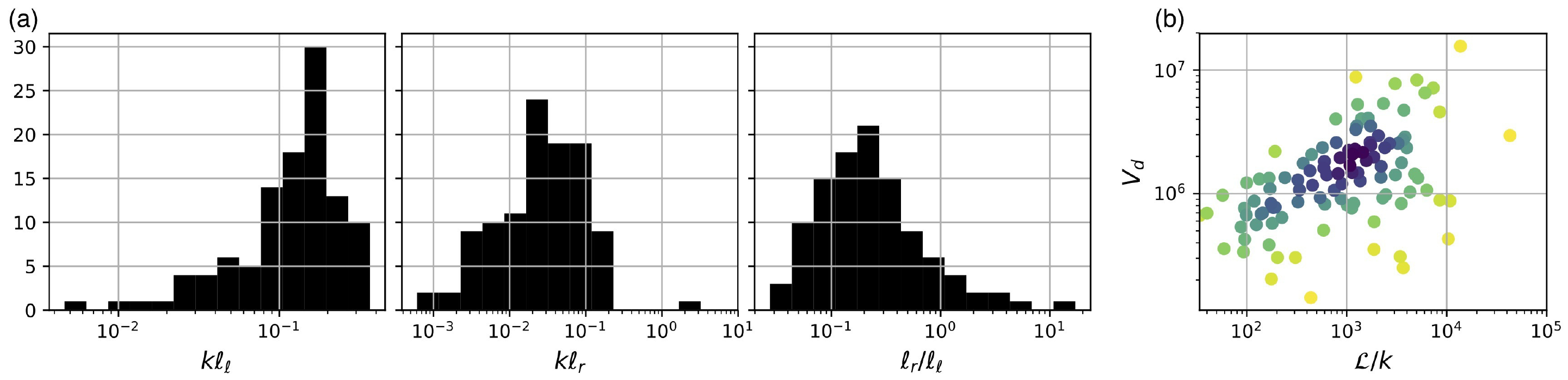}
        \caption{(a) Histograms of the fitted parameters  $k\ell_\ell$, $k\ell_\textrm{r}$ and $\ell_\textrm{r}/\ell_\ell$ for the US stocks presented in Table~\ref{tablestocks}. {(b) Average daily traded volume $V_\textrm{d}$ as function of the liquidity parameter $\mathcal L/k$. The data points are coloured by density (from yellow to dark blue).}}
\label{imm:histograms}
\end{figure*}

The fitted values of $k\ell_\ell$ and  $k\ell_\textrm{r}$ for the Euro Stoxx contract and the US stocks presented in Table~\ref{tablestocks}  are reported in Figs.~\ref{imm:stab}(a) and (b).
Figure~\ref{imm:histograms}(a) displays the statistics of $k\ell_\ell$, $k\ell_\textrm{r}$ and $\ell_\textrm{r}/\ell_\ell$ for the stocks. As one can see the values of $k\ell_\ell$ stand below the critical line by typically one order of magnitude. Also most of the data is consistent with $\ell_\textrm{r} < \ell_\ell$. More precisely, the ratio $\ell_\textrm{r}/\ell_\ell$ is typically $\approx 3-10$, which is ${D_\textrm{r}}/D_\ell$ typically $\approx 0.01-0.1$, consistent with our initial intuition that diffusivity is much smaller in the revealed order book than in its latent counterpart. While not a strong effect, the $\ell_\textrm{r}/\ell_\ell$ ratios are on average slightly smaller for the small tick assets (Fig.~\ref{imm:stab}(a)). The daily traded volume $V_\textrm{d}$ does not seem to play a notable role in the distribution of the data in the $(k\ell_\textrm{r} ,k\ell_\ell)$ map (Fig.~\ref{imm:stab}(b)).  {However, $\mathcal L/k$ shows positive correlation with $V_\textrm{d}$ (see Fig \ref{imm:histograms}(b)), consistent with the interpretation given to the liquidity parameter $\mathcal L$ in the latent order book models \cite{donierLLOB}. {In order to evaluate the potential relevance of our model to quantitatively account for market stability, we have calibrated the Eurostoxx data on a particularly agitated subset, that is the two hours about the flash crash of February 5th 2018 at 8:15pm (Paris time), see light blue star on pannels (a) and (b) of Fig.~\ref{imm:stab}. One finds, as expected, that the new data point is closer to the critical line than the reference point (white star), indicating lesser stability during the flash crash period.}}\\

\section{Price impact}
\label{sec:impact}

In this section we study how liquidity reacts, in our model, to the   presence of a metaorder, namely a large trading order split into small orders executed incrementally. Following Donier \textit{et al.} \cite{donierLLOB} we introduce a metaorder as an additional  current of buy/sell particles falling precisely at the trade price. The system of equations governing the system in the presence of a metaorder is left unchanged for  the latent  order book (Eqs.~\eqref{lorenzo1} and \eqref{lorenzo2}), while the RHS of Eq.~\eqref{lorenzo3} must be complemented with the extra additive term $+\ m_t \delta(x-p_{t})$,   representing the metaorder, with  $m_t$ the execution rate {and $\delta$ the Dirac delta function}. In the following we restrict to buy metaorders with constant execution rates $m_t = m_0 > 0$, without loss of generality since the sell metaorder $m_0 < 0$ is perfectly symmetric. In order to extract the dimensionless parameters governing the dynamic system, we introduce  $\tilde{\xi} = k\xi, \tilde{t} = \omega t, \tilde{\rho} = {k}\rho/{\mathcal{L}} , \tilde{\phi}_\textrm{r} = {k}\phi_\textrm{r}/{\mathcal{L}} $
and write the equations in a dimensionless form:
\begin{subeqnarray}
\partial_{\tilde{t}} \tilde{\rho}_\textrm{B}^{(\ell)}  &=&( k \ell_{\ell})^2\partial_{\tilde{x}\tilde{x}} \tilde{\rho}_\textrm{B}^{(\ell)}  - \left\{\Gamma(-\tilde{\xi})\tilde{\rho}_\textrm{B}^{(\ell)} -[1-\Gamma(-\tilde{\xi})] \mathds{1}_{\{\tilde{x} < \tilde{p}_{\tilde{t}} \}} \tilde{\phi}_\textrm{r} \right\} \quad \\
\partial_{\tilde{t}} \tilde{\rho}_\textrm{A}^{(\ell)}  &=& ( k \ell_{\ell})^2\partial_{\tilde{x}\tilde{x}} \tilde{\rho}_\textrm{A}^{(\ell)}  - \left\{\Gamma(\tilde{\xi})\tilde{\rho}_\textrm{A}^{(\ell)} +[1-\Gamma(\tilde{\xi})] \mathds{1}_{\{\tilde{x} > \tilde{p}_{\tilde{t}} \}}  \tilde{\phi}_\textrm{r} \right\} \quad \\
\partial_{\tilde{t}}  \tilde{\phi}_\textrm{r}  &=& ( k \ell_{\textrm{r}})^2\partial_{\tilde{x}\tilde{x}}  \tilde{\phi}_\textrm{r}  
  - \left\{\Gamma(\tilde{\xi})\tilde{\rho}_\textrm{A}^{(\ell)}  - \Gamma(-\tilde{\xi})\tilde{\rho}_\textrm{B}^{(\ell)} +[1-\Gamma(|\tilde{\xi}|)] \tilde{\phi}_\textrm{r} \right\} + (m_0/\mathcal J)   \delta(\tilde{\xi}) \ , \slabel{lasteq}
\label{eq:general system metaorder adimensional}
\end{subeqnarray}
 with  $\mathcal J = \mathcal L\omega/k^2$ the typical scale of the rate at which latent orders are revealed (recall that $\mathcal L/k^2$ is the typical available volume in the latent order book that have a substantial probability to be revealed). Matching the first and third terms on the right hand side of Eq.~\eqref{lasteq} yields a relevant dimensionless number $(m_0/\mathcal J)/( k \ell_{\textrm{r}})^2 = m_0/{J_\textrm{r}}$ with $J_\textrm{r} = D_\textrm{r}\mathcal L$. In the case $\ell_\textrm{r}=0$ the relevant dimensionless number is simply given by $m_0/\mathcal{J}$. 
 \\

\begin{figure}[t!]
    \centering
        \includegraphics[width=1\columnwidth]{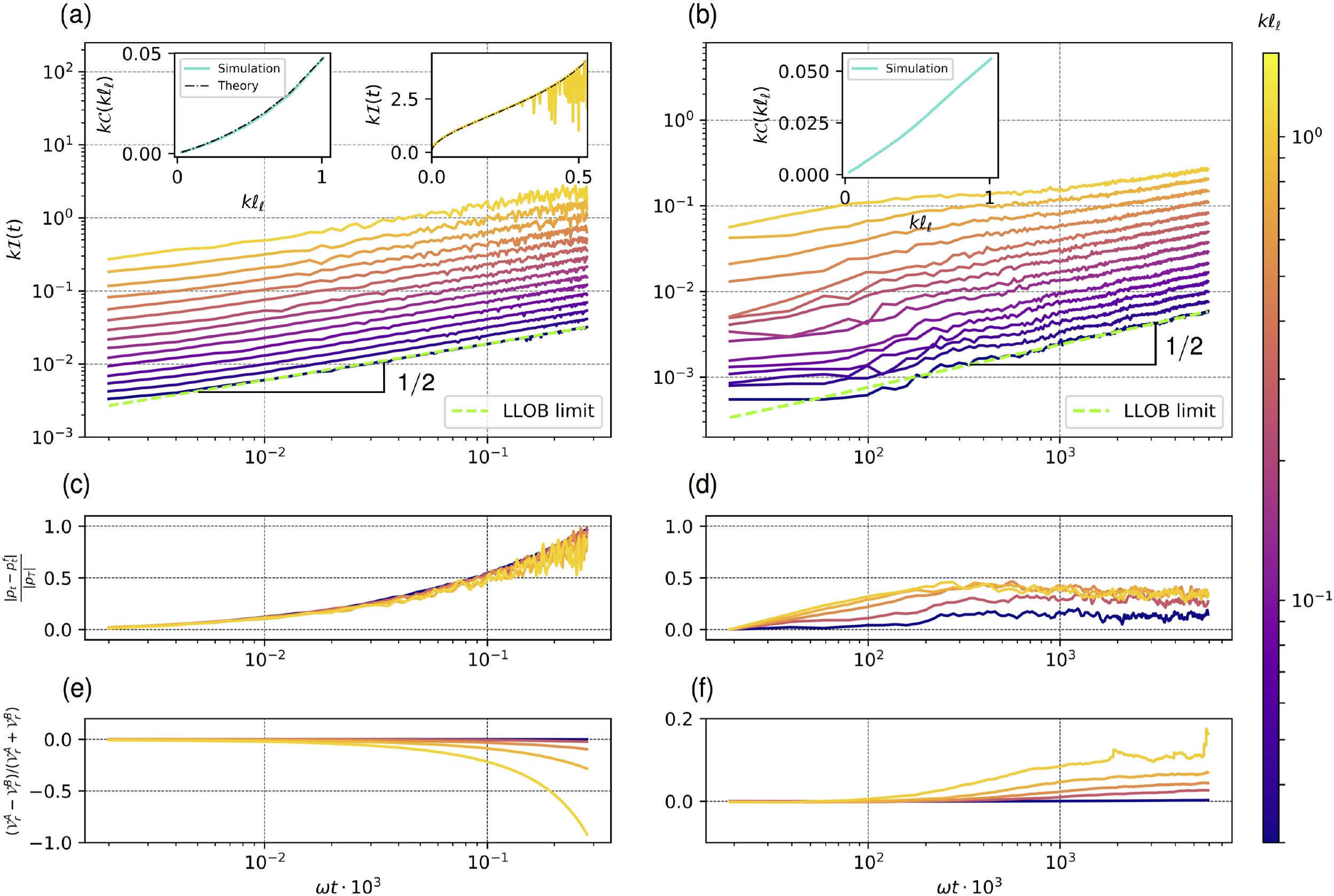}
    \caption{Price impact for $\ell_\textrm{r}=\ell_\ell$.  (a)  $m_0\gg{J_\textrm{r}}$ (b)  $m_0\ll{J_\textrm{r}}$. The dashed green lines indicate the LLOB limits,  {$I_\textrm{LLOB}(t) = \sqrt{ \alpha {Q_t}/(\pi \mathcal{L})}$} {with $\alpha = m_0/{J_u}$} for the slow regime and  $I_\textrm{LLOB}(t) = \sqrt{{2{Q_t}}/{\mathcal{L}}}$ for the fast regime. {The top left insets on each plot indicate the factor $c$ as function of ${k}{\ell_\ell}$ defined as $I(t) = c\,I_\textrm{LLOB}(t)$. 
}The top right inset of subplot (a) shows an extreme regime with very high execution rate, the dash-dotted line indicates the theoretical prediction as given by the numerical inversion of Eq.~\eqref{eq:geometrical impact DrDU}.  Subplots (c) and (d) display relative price difference between the trade price and the fair price (see Eq.~\eqref{eq: fair price}). Subplots (e) and (f) display the relative revealed volume imbalance. {Typical computational time: a few \,hours to obtain all the impact curves.}}
\label{imm:impact1}
\end{figure}

In order to compute the price impact $I(Q_t)=\mathbb E[p_t-p_0|Q_t=m_0t]$, we performed numerical simulations of our model in the presence of a metaorder in several limit cases. {More specifically, starting from the stationary order books, we implemented a buy/sell metaorder as an additional current of buy/sell particles falling precisely at the best ask/bid, thus acting as a sequence of market orders. We stored the order books as a function of time to be able to compute dynamically the relevant scalar quantities below.} We explored in particular $\ell_\textrm{r}=\ell_\ell$ and $\ell_\textrm{r}=0$, in both high and low participation rate regimes, for different values of $k\ell_\ell$ (see Figs.~\ref{imm:impact1} and \ref{imm:impact2}). \\

Before presenting the results of the numerical simulations, note that there exists a regime where the calculations can be brought a little bit further analytically, that is when we can give a geometrical interpretation to the problem. When the book is almost static on the time scale of the metaorder execution \textit{i.e.}  $m_0 \gg J_\textrm{r}$ (resp.   $m_0 \gg \mathcal J$ for $\ell_\textrm{r}=0$){, one has} $
\int_{0}^{t} \textrm ds \,m_0 = -\int_{0}^{p_\textrm{t}} \textrm d\xi \phi_\textrm{r}^{\textrm{st}}(\xi) $ with $\xi = x - p_0$. In particular for $\ell_\textrm{r} = 0$ one obtains:
\begin{eqnarray}
Q_t &=& -\frac{\mathcal{L}\ell_\ell^2}{2}\left(1-e^{-p_\textrm{t}/\ell_\ell}\right) + \frac{\mathcal{L}p_\textrm{t}}{k}\log\left(1-e^{-kp_\textrm{t}}\right) - \frac{\mathcal{L}}{k^2}\left[\textrm{Li}_2(e^{-kp_\textrm{t}}) - \textrm{Li}_2(1)\right] \ ,
\label{eq:geometrical impact Dr0}
\end{eqnarray}
where $\textrm{Li}_2(y)=\sum_{k=1}^\infty y^k/k^2$ stands for the polylogarithm of order 2, and can be inverted numerically to obtain the price trajectory $p_t$. Similarly, for $\ell_\textrm{r} = \ell_\ell$ one has:
\begin{eqnarray}
Q_t &=& \frac{\mathcal{L}\alpha}{k^2}\left\{[k(p_t+\beta)+1]e^{-kp_t}-(k\beta+1)\right\} - \frac{\mathcal{L}\ell_\ell\gamma}{1+k\ell_\ell}\left(1-e^{-(k+1/\ell_\ell)p_t}\right)  \nonumber \\
&&+\ \mathcal{L}\ell_\ell{\eta} p_t e^{-p_t/\ell_\ell} - \mathcal{L}[{\eta}\ell_\ell^2 - \ell_\ell(\alpha \beta+\gamma)]\left(1-e^{-p_t/\ell_\ell}\right) \ ,
\label{eq:geometrical impact DrDU}
\end{eqnarray}
where $\alpha = [(k\ell_\ell)^2-1]^{-1}$, $\beta=2\alpha k\ell_\ell^2$, $\gamma = g(k\ell_\ell)[k^2\ell_\ell(k\ell_\ell+2)]^{-1}$, ${\eta} = g(k\ell_\ell)[2k\ell_\ell]^{-1}$. 
In the following we discuss the more general numerical results for both limit cases $\ell_\ell = \ell_\textrm{r}$ and $\ell_\textrm{r}=0$.

\subsubsection*{The case $\boldmath\ell_\ell = \ell_\textrm{r}$}
The main plots in Figs.~\ref{imm:impact1}(a) and (b) display robust square root price trajectories, regardless of the values of $k \ell_\ell$. For $m_0 \gg J_\textrm{r}$ the price trajectory matches the theoretical prediction  given above inverting Eq.~\eqref{eq:geometrical impact DrDU}. As expected from the exponentially vanishing liquidity when $x-p_0>k^{-1}$, impact eventually diverges for very extreme regimes, see top right inset in Fig.~\ref{imm:impact1}(a). 
For $k \ell_\ell \ll 1$, one recovers the LLOB limit  in both fast and slow regimes, also as expected. For non vanishing values  of $k \ell_\ell$, the  impact increases with increasing $k \ell_\ell$. In particular for $m_0 \gg J_\textrm{r}$ one obtains at short times  (equivalently small volumes):
\begin{equation}
p_\textrm{t} = \sqrt{\frac{2{Q_t}}{|\partial_x \phi_\textrm{r}(0^+)|}} \sim  \sqrt{2{Q_t}} \left(1-\frac{k \ell_\ell}{\zeta_\textrm{c}} \right)^{\hspace{-1mm}-{1}/{2}}.
\label{eq:impact DrDu fast}
\end{equation}
As expected, impact diverges when at the incipient liquidity crisis point. Figures~\ref{imm:impact1}(c) and (d) display the relative distance between the trade price and the fair price as function of time. In the fast execution regime all curves fall on top of each other and $|p_t-p_t^\textrm{f}|\approx {|p_t|}$, consistent with the idea that the book (in particular latent) does not have time to reassess during the execution, and as a consequence the fair price varies at a much slower rate than the trade price. A different scenario takes place in the small execution rate regime. We observe that the relative distance between trade and fair prices stabilizes. In other terms, the latent order book evolves at a speed that is comparable to that of the metaorder and the fair price follows the trade price quite accurately.  
 \\

The evolution of relative volume imbalance (see Figs.~\ref{imm:impact1}(e) and (f)) allows one to draw similar conclusions. In the fast execution limit, the imbalance diverges as the latent order book has no time to refill the revealed order book (this effect is all the more evident as we approach the  vanishing liquidity limit $k \ell_\ell\to \zeta_\textrm{c}$), while in the slow execution limit the imbalance is much smaller. Most importantly, note that in this limit the imbalance becomes positive. This is consistent with the fact that when the trade price moves slowly, the conversion probability $\Gamma$ is shifted with it and new orders  reveal on top of the existing ones to supply the metaorder, while the orders left behind progressively unreveal. 

\subsubsection*{The case $\boldmath\ell_\textrm{r} = 0$}

\begin{figure}[t!]
    \centering
        \includegraphics[width=1\columnwidth]{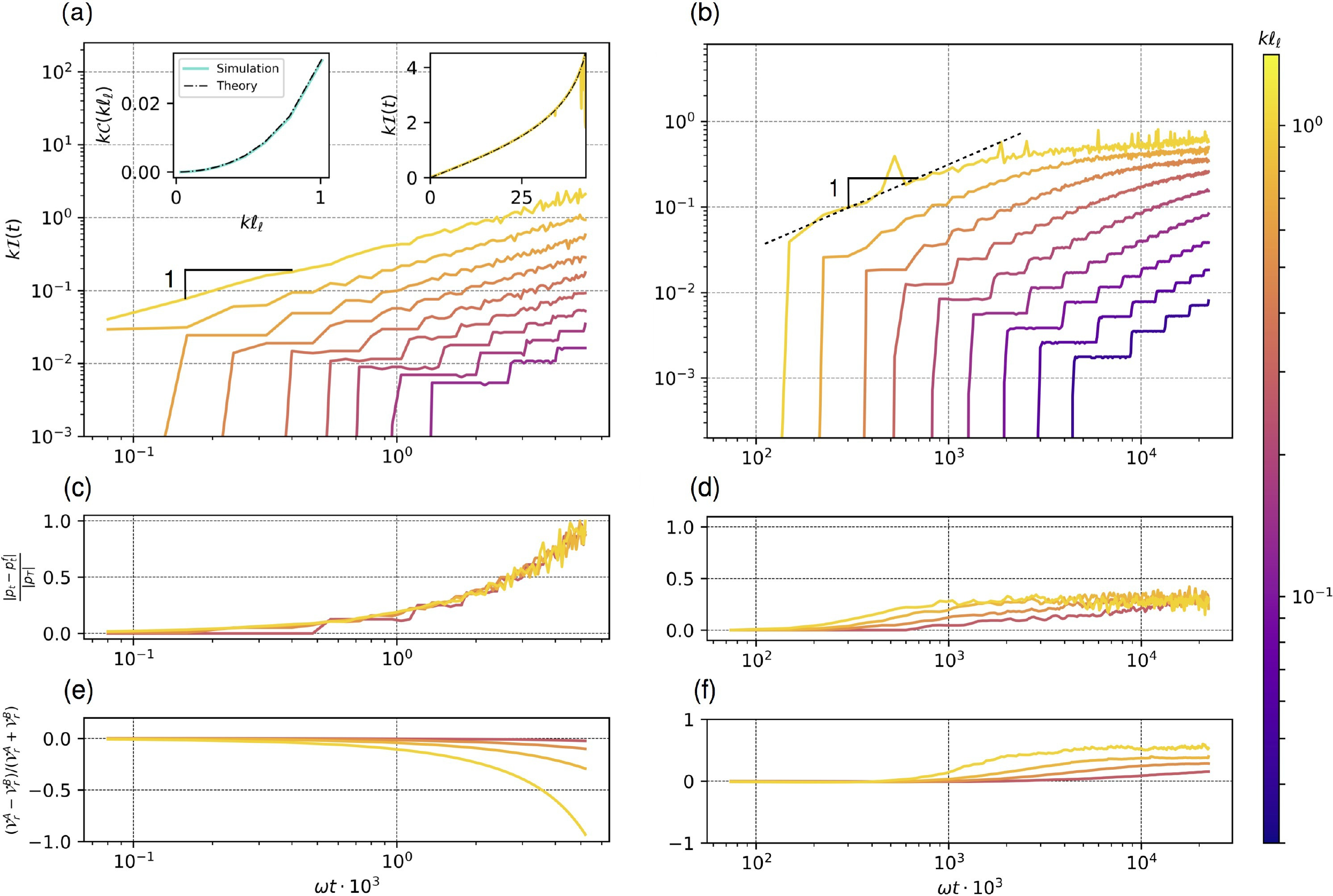}
    \caption{Price impact for $\ell_\textrm{r}=0$. (a)  and $m_0 \gg \mathcal J $ (b)  $m_0\ll\mathcal J$. The top left insets on subplot (a) indicates the factor $c$ as function of ${k}{\ell_\ell}$ here defined as ${I}(t) = {c}t$ for $\ell_\textrm{r}=0$. The top right insets show an extreme regime with very high execution rate for which the impact diverges. Subplots (c) and (d) display relative price difference between the trade price and the fair price, see Eq.~\eqref{eq: fair price}. Subplots (e) and (f) display the relative revealed volume imbalance. {Typical computational time: a few hours to obtain all the impact curves.}}
\label{imm:impact2}
\end{figure}

The limit $\ell_\textrm{r} \to 0$ corresponds to $m_0/J_\textrm{r} \to \infty$, so in some sense one could say that we are always in a high participation rate regime. However the absence of diffusion means that the system can only evolve through the revealing and unrevealing currents. As mentioned above, in this case the relevant dimensionless number becomes $m_0/\mathcal{J}$, that shall be referred to as participation rate in the following.\\ 
 
Figure~\ref{imm:impact2}(a) displays price trajectories in the fast execution regime, here $m_0\gg \mathcal J$. The metaorder is faster than the revealing current and the price trajectory is given by inverting Eq.~\eqref{eq:geometrical impact Dr0}. At short times,  one obtains (see Eq.~\eqref{eq: scaling Dr0}):
\begin{equation}
p_\textrm{t} = \frac{k{Q_t}}{\mathcal{L}}\left(1-\frac{k \ell_\ell}{\zeta_\textrm{c}}\right)^{-1} \ ,
\label{eq:impact Dr0 fast}
\end{equation}
which is  linear impact, consistent with $\phi_\textrm{r}(0) \neq 0$. Analogous to the case $\ell_\ell = \ell_\textrm{r}$, in extreme regimes the impact eventually diverges (see top right inset of Fig.~\ref{imm:impact2}(a)). Regarding the imbalance and the relative distance between the trade price and the fair price, the interpretation is analogous to the $\ell_\textrm{r} = \ell_\ell$ case. \\

Figure~\ref{imm:impact2}(b) displays the price trajectories in the low participation rate regime $m_0\ll \mathcal J$. Here
 the impact is genuinely linear at short times but crossovers to concave after a typical time $t^\star\sim \mathcal V_\textrm{r}/m_0$ with $\mathcal V_\textrm{r}$ the typical volume available in the revealed order book. This interesting regime can be easily  understood as follows.  At short times the metaorder is executed against the orders present in the stationary locally constant revealed order book (linear impact), but after a while liquidity revealing from the linear latent order book takes over (implying concave impact).\footnote{Note that recovering precisely the square root law in this regime is quite challenging because the numerical simulation, by essence discontinuous, induces artificial spread effects: the spread widens and does not get refilled as fast as it should, due to limited resolution and vanishing liquidity in the price region. Obtaining smooth results requires a lot of averaging.} More precisely, as $k\ell_\ell$ is increased, $t^\star$ is decreased consistent with the idea that: the larger $k\ell_\ell$, the smaller the revealed liquidity $\mathcal V_\textrm{r}$ and thus the sooner it gets consumed. Note that linear impact at short times (equiv. small volumes) has been reported empirically in the literature \cite{Zarinelli}. An alternative theoretical framework to understand linear impact at short times is provided in \cite{benzaquen2018market} with the introduction of two types of agents, fast and slow. Our model, in its final version, should also include the possibility for agents with different timescales, possibly with different conversion rates $\omega$.
In Fig.~\ref{imm:impact2}(d), we observe that the relative distance between trade and fair prices stabilizes after the typical time $t^\star$. In other terms, for $t>t^\star$ the latent order book evolves at a rate  comparable to that of the metaorder, and the fair price  follows the trade price with some constant lag. Figure~\ref{imm:impact2}(f) displays similar conclusions.

\section{Concluding remarks}
\label{sec:ccl}

We have proposed a simple, consistent model describing how latent liquidity gets revealed in the real order book. As a first step in the rapprochement of latent order book models and real order book data, our study upgrades the former from a \emph{toy model} status to an observable setup that can be quantitatively calibrated on real data.
 Although probably too simple in its present form, our main contribution is to show how the latent order book could be inferred from its revealed counterpart. One of our key theoretical result is the existence of a market instability threshold, where the conversion of the trading intentions of market participants into \emph{bona fide} limit orders becomes too slow, inducing \emph{liquidity crises}. From a regulatory perspective, our model indicates how assets can be sorted according to their position in the stability map, a proxy for their propensity to liquidity dry-ups and large price jumps. In particular, our setup could constitute an insightful alternative to relate stability and tick size, a subject that has raised the attention of many in the recent past (see e.g. \cite{dayri2015large}). {To confirm potential perspectives for market regulation, we showed that our model exhibits lesser stability when calibrating the same asset during a period of higher volatility}.
\\

\begin{figure*}[t!]
        \centering
        \includegraphics[width=0.99\columnwidth]{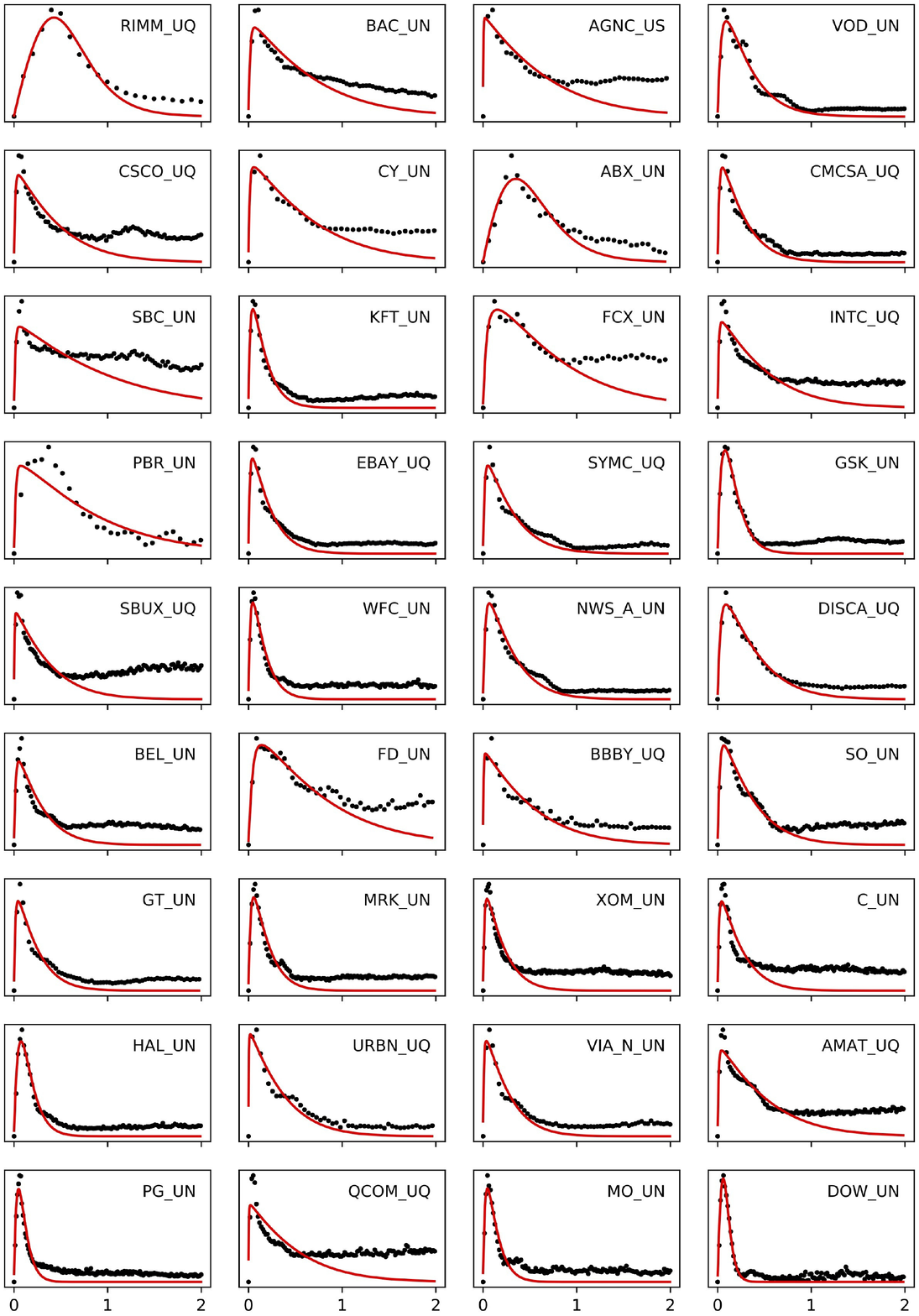}
        \caption{Fit of the stationary revealed order book $\phi_\textrm{r}$ to the average order books of over one hundred US stocks, see Table~\ref{tablestocks}.}
\label{imm:fit_stoks1}
\end{figure*}

\begin{figure*}[t!]
        \centering
        \includegraphics[width=0.99\columnwidth]{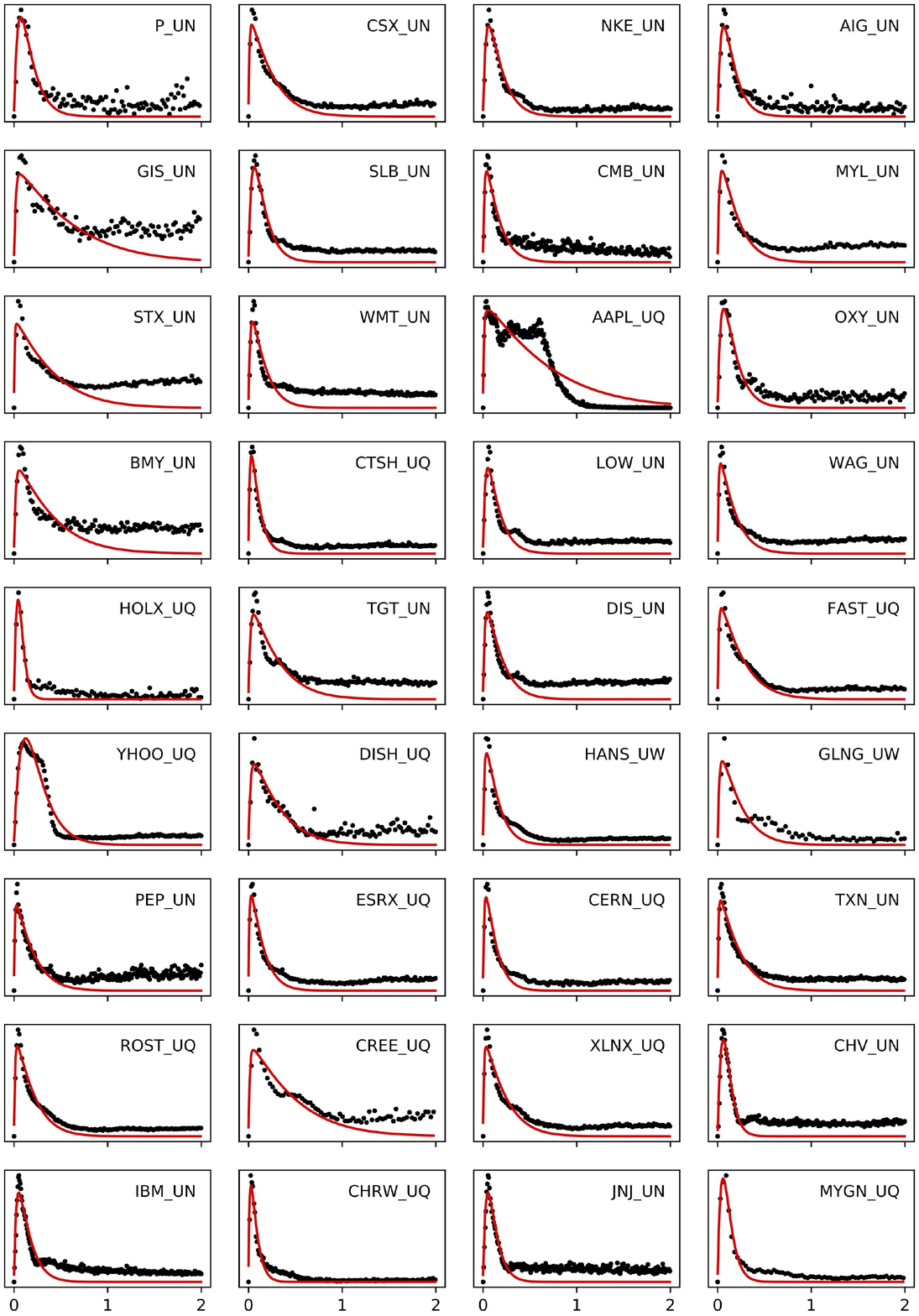}
        \caption{Fit of the stationary revealed order book $\phi_\textrm{r}$ to the average order books of over one hundred US stocks, see Table~\ref{tablestocks}.}
\label{imm:fit_stoks2}
\end{figure*}

\begin{figure*}[t!]
        \centering
        \includegraphics[width=0.99\columnwidth]{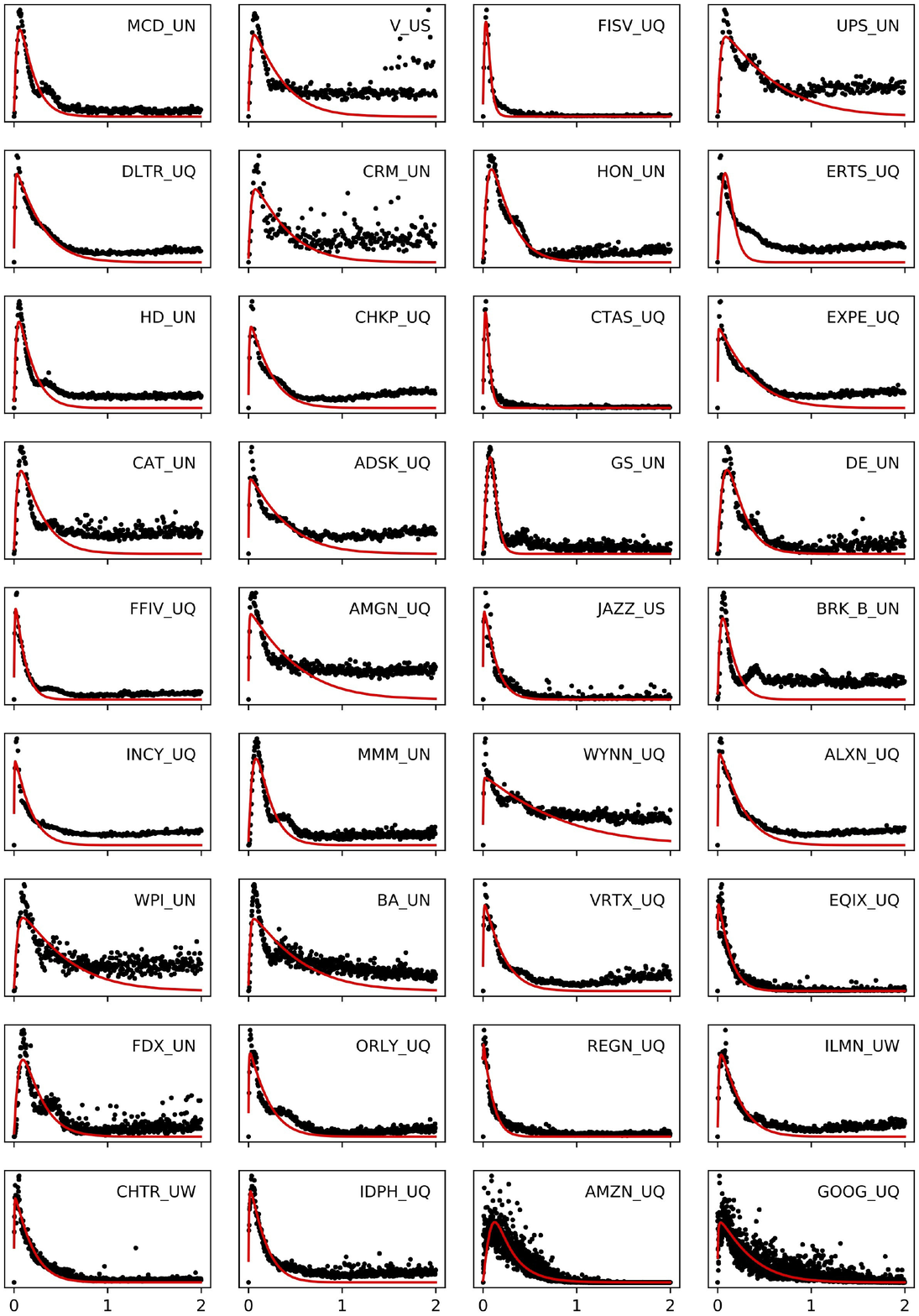}
        \caption{Fit of the stationary revealed order book $\phi_\textrm{r}$ to the average order books of over one hundred US stocks, see Table~\ref{tablestocks}.}
\label{imm:fit_stoks3}
\end{figure*}

While providing quite satisfactory results on several grounds, our model lacks a number of features that must be addressed in future work. In particular, being an inheritance of the LLOB model \cite{donierLLOB}, our model does not solve the so-called diffusivity puzzle (even persistent order flow lead to mean-reverting prices -- see \cite{mastromatteo2014agent,benzaquen2018fractional,benzaquen2018market}). An interesting extension is suggested by the stocks data presented in Figs~\ref{imm:fit_stoks1}, \ref{imm:fit_stoks2} and \ref{imm:fit_stoks3}. Indeed, some of the order books display a bimodal shape indicating that they would be better fitted with a model involving two conversion timescales (equivalently two price scales). The recent multi-timescale liquidity setup  (see \cite{benzaquen2018market}) allows for both a \emph{fast} liquidity of high frequency market makers, and a \emph{slow} liquidity of directional traders, while resolving several difficulties of latent order book models. Such a framework should output bimodal distributions and yield even better fits of the real order book, allowing one to infer different liquidity timescales.  \\

We thank Fr\'ed\'eric Bucci, Alexandre Darmon, Jonathan Donier, Stephen Hardiman and Iacopo Mastromatteo for fruitful discussions.

\bibliographystyle{iopart-num}
\bibliography{bibs}


\clearpage

\begin{changemargin}{-1cm}{-1cm}

{\scriptsize 

\begin{multicols}{2}

\setlength{\tabcolsep}{2.5pt}

\noindent \begin{tabular}{lrrrrrrr}
\toprule
Stock &    price $\$$&  S  \ \  & $10^{-6}V_\textrm{d}$ &    $ 10^{-3} \mathcal L$ &     $ k $ \ \  &    $  \ell_\ell $ \ \ &     $ \ell_\textrm{r}$ \ \  \quad \\
\midrule
RIMM     &    10.03 &   1.003 &    0.87 &  602.0 &   4.71 &  0.0020 &  0.0059 \\
BAC      &    28.66 &   1.007 &   15.62 &   34.3 &   2.50 &  0.0350 &  0.0033 \\
AGNC     &    19.92 &   1.009 &    1.03 &   10.8 &   2.51 &  0.0165 &  0.0005 \\
VOD      &    29.62 &   1.011 &    0.89 &   47.6 &   5.58 &  0.0230 &  0.0090 \\
CSCO     &    38.57 &   1.012 &    7.17 &   26.7 &   3.64 &  0.0309 &  0.0027 \\
CY       &    16.14 &   1.018 &    1.44 &   11.9 &   2.49 &  0.0195 &  0.0020 \\
ABX      &    16.46 &   1.023 &    2.95 &  191.1 &   4.45 &  0.0031 &  0.0060 \\
CMCSA    &    37.65 &   1.023 &    6.54 &   38.3 &   6.31 &  0.0230 &  0.0046 \\
SBC      &    36.51 &   1.029 &    7.76 &    5.3 &   1.75 &  0.0417 &  0.0023 \\
KFT      &    42.25 &   1.036 &    2.88 &   39.1 &  10.22 &  0.0131 &  0.0047 \\
FCX      &    16.45 &   1.040 &    4.74 &    8.3 &   2.22 &  0.0201 &  0.0054 \\
INTC     &    44.87 &   1.042 &    8.31 &   16.7 &   3.31 &  0.0318 &  0.0024 \\
PBR      &    13.54 &   1.044 &    4.59 &   16.9 &   1.99 &  0.1284 &  0.0035 \\
EBAY     &    39.12 &   1.052 &    2.73 &   28.1 &   7.71 &  0.0176 &  0.0037 \\
SYMC     &    28.96 &   1.072 &    1.78 &   19.4 &   5.50 &  0.0227 &  0.0040 \\
GSK      &    37.99 &   1.074 &    0.83 &   38.4 &  10.94 &  0.0084 &  0.0069 \\
SBUX     &    57.01 &   1.076 &    2.95 &    9.4 &   4.53 &  0.0178 &  0.0009 \\
WFC      &    56.31 &   1.082 &    5.37 &   25.6 &  11.02 &  0.0156 &  0.0057 \\
NWS      &    33.21 &   1.100 &    2.55 &   18.7 &   5.74 &  0.0262 &  0.0065 \\
DISCA    &    22.57 &   1.101 &    1.42 &   13.7 &   4.46 &  0.0247 &  0.0072 \\
BEL      &    49.35 &   1.104 &    4.09 &   10.0 &   6.11 &  0.0278 &  0.0048 \\
FD       &    23.44 &   1.117 &    2.46 &    3.9 &   2.26 &  0.0411 &  0.0093 \\
BBBY     &    22.20 &   1.118 &    1.35 &    7.0 &   3.17 &  0.0178 &  0.0008 \\
SO       &    44.21 &   1.143 &    1.69 &    5.2 &   4.87 &  0.0323 &  0.0060 \\
GT       &    31.36 &   1.191 &    0.98 &   17.8 &   7.24 &  0.0169 &  0.0035 \\
MRK      &    56.76 &   1.210 &    3.54 &   11.5 &   9.16 &  0.0209 &  0.0071 \\
XOM      &    80.51 &   1.219 &    4.03 &   11.2 &   8.03 &  0.0231 &  0.0042 \\
C        &    73.26 &   1.225 &    5.28 &    8.8 &   6.83 &  0.0261 &  0.0040 \\
HAL      &    47.17 &   1.254 &    2.35 &   28.7 &  11.94 &  0.0117 &  0.0097 \\
URBN     &    23.44 &   1.257 &    0.92 &   10.5 &   4.57 &  0.0123 &  0.0007 \\
VIA      &    29.54 &   1.264 &    1.66 &   14.2 &   6.45 &  0.0127 &  0.0020 \\
AMAT     &    53.43 &   1.315 &    3.53 &    5.6 &   3.23 &  0.0305 &  0.0021 \\
PG       &    84.03 &   1.401 &    2.60 &   30.7 &  17.95 &  0.0098 &  0.0078 \\
QCOM     &    59.38 &   1.416 &    3.30 &    4.0 &   3.25 &  0.0177 &  0.0006 \\
MO       &    64.07 &   1.428 &    2.09 &   14.9 &  13.02 &  0.0143 &  0.0064 \\
DOW      &    64.99 &   1.439 &    2.35 &  100.6 &  25.26 &  0.0015 &  0.0030 \\
P        &    52.99 &   1.439 &    1.85 &   17.5 &  11.15 &  0.0149 &  0.0104 \\
CSX      &    54.49 &   1.561 &    2.16 &    9.8 &   6.77 &  0.0199 &  0.0026 \\
NKE      &    63.01 &   1.586 &    2.29 &   12.4 &  10.24 &  0.0191 &  0.0088 \\
AIG      &    61.12 &   1.592 &    1.48 &   15.9 &  12.27 &  0.0168 &  0.0109 \\
GIS      &    49.88 &   1.645 &    1.17 &    1.3 &   2.79 &  0.0431 &  0.0037 \\
SLB      &    67.36 &   1.661 &    2.20 &   14.9 &  11.92 &  0.0167 &  0.0097 \\
CMB      &   106.81 &   1.667 &    4.03 &    8.6 &  11.14 &  0.0199 &  0.0048 \\
MYL      &    39.10 &   1.748 &    1.98 &   16.1 &   8.65 &  0.0184 &  0.0049 \\
STX      &    45.28 &   1.822 &    1.48 &    4.6 &   4.09 &  0.0212 &  0.0013 \\
WMT      &    90.73 &   1.884 &    2.59 &    8.3 &  10.61 &  0.0211 &  0.0055 \\
AAPL     &   168.14 &   1.922 &    8.78 &    3.1 &   2.51 &  0.0107 &  0.0011 \\
OXY      &    64.65 &   1.971 &    1.20 &    9.6 &  10.77 &  0.0191 &  0.0101 \\
BMY      &    61.12 &   2.032 &    1.76 &    1.5 &   4.21 &  0.0372 &  0.0044 \\
CTSH     &    75.94 &   2.129 &    1.27 &   23.7 &  17.41 &  0.0095 &  0.0040 \\
LOW      &    81.69 &   2.204 &    1.61 &    6.9 &  11.71 &  0.0194 &  0.0078 \\
WAG      &    71.64 &   2.298 &    2.25 &   10.3 &   9.79 &  0.0176 &  0.0032 \\
HOLX     &    42.09 &   2.314 &    0.88 &  317.2 &  29.35 &  0.0008 &  0.0012 \\
TGT      &    66.76 &   2.409 &    1.53 &    2.4 &   5.62 &  0.0320 &  0.0050 \\
DIS      &   103.38 &   2.440 &    2.36 &    5.4 &   9.51 &  0.0236 &  0.0057 \\
FAST     &    50.13 &   2.495 &    0.91 &    5.9 &   6.68 &  0.0213 &  0.0032 \\
YHOO     &    71.07 &   2.649 &    2.55 &   21.6 &   8.05 &  0.0042 &  0.0043 \\
DISH     &    48.54 &   2.671 &    0.82 &    3.3 &   5.53 &  0.0309 &  0.0059 \\
HANS     &    59.37 &   2.751 &    0.84 &   13.8 &  11.79 &  0.0163 &  0.0047 \\
GLNG     &    27.57 &   2.925 &    0.35 &   14.1 &   7.52 &  0.0187 &  0.0048 \\
PEP      &   112.52 &   3.009 &    1.42 &    5.3 &   8.42 &  0.0174 &  0.0020 \\
ESRX     &    69.15 &   3.032 &    1.45 &    9.6 &  11.69 &  0.0137 &  0.0029 \\
CERN     &    67.81 &   3.100 &    0.81 &   13.6 &  13.42 &  0.0134 &  0.0037 \\
TXN      &    99.90 &   3.131 &    1.82 &    4.6 &   7.61 &  0.0200 &  0.0025 \\\bottomrule
\end{tabular}

\begin{tabular}{lrrrrrrr}
\toprule
Stock &    price $\$$&  S  \ \  & $10^{-6}V_\textrm{d}$ &    $ 10^{-3} \mathcal L$ &     $ k $ \ \  &    $  \ell_\ell $ \ \ &     $ \ell_\textrm{r}$ \ \  \quad \\
\midrule
ROST     &    70.66 &   3.149 &    1.06 &    6.2 &   8.32 &  0.0198 &  0.0033 \\
CREE     &    34.92 &   3.165 &    0.51 &    2.0 &   3.48 &  0.0322 &  0.0030 \\
XLNX     &    70.21 &   3.199 &    0.93 &    4.0 &   7.34 &  0.0200 &  0.0026 \\
CHV      &   116.12 &   3.867 &    1.95 &   16.8 &  19.58 &  0.0085 &  0.0108 \\
IBM      &   152.11 &   3.917 &    1.29 &    3.9 &  11.96 &  0.0218 &  0.0083 \\
CHRW     &    87.75 &   3.980 &    0.59 &   41.7 &  22.18 &  0.0075 &  0.0034 \\
JNJ      &   131.86 &   4.007 &    2.07 &    6.9 &  15.56 &  0.0149 &  0.0098 \\
MYGN     &    33.28 &   4.032 &    0.31 &   61.1 &  17.84 &  0.0019 &  0.0021 \\
MCD      &   156.99 &   4.131 &    1.07 &    3.5 &  10.38 &  0.0276 &  0.0109 \\
V        &   121.14 &   4.170 &    2.20 &    1.0 &   5.21 &  0.0381 &  0.0058 \\
FISV     &   106.11 &   4.408 &    0.43 &  377.8 &  36.42 &  0.0008 &  0.0010 \\
UPS      &   107.88 &   5.055 &    0.87 &    0.4 &   3.02 &  0.0566 &  0.0071 \\
DLTR     &    97.27 &   5.124 &    0.86 &    2.0 &   6.07 &  0.0206 &  0.0019 \\
CRM      &   116.98 &   5.563 &    1.22 &    0.5 &   4.61 &  0.0452 &  0.0074 \\
HON      &   148.04 &   5.754 &    0.78 &    1.4 &   7.36 &  0.0365 &  0.0151 \\
ERTS     &   116.94 &   5.872 &    1.07 &  220.6 &  34.96 &  0.0101 &  0.0641 \\
HD       &   175.53 &   5.944 &    1.35 &    2.4 &   9.83 &  0.0275 &  0.0079 \\
CHKP     &   104.95 &   6.161 &    0.38 &    1.3 &   7.57 &  0.0183 &  0.0020 \\
CTAS     &   161.47 &   6.443 &    0.25 &  141.6 &  38.58 &  0.0038 &  0.0040 \\
EXPE     &   124.49 &   6.716 &    0.83 &    0.8 &   4.82 &  0.0114 &  0.0006 \\
CAT      &   153.78 &   6.934 &    1.10 &    1.2 &   6.78 &  0.0408 &  0.0106 \\
ADSK     &   118.11 &   6.967 &    0.77 &    0.8 &   4.34 &  0.0181 &  0.0010 \\
GS       &   230.09 &   6.976 &    0.76 &   30.5 &  27.22 &  0.0011 &  0.0038 \\
DE       &   154.44 &   6.985 &    0.58 &    1.6 &   8.70 &  0.0338 &  0.0216 \\
FFIV     &   128.76 &   7.457 &    0.30 &    4.9 &  15.98 &  0.0095 &  0.0016 \\
AMGN     &   179.67 &   7.504 &    1.31 &    0.5 &   3.42 &  0.0187 &  0.0009 \\
JAZZ     &   146.67 &   8.266 &    0.14 &    5.6 &  12.72 &  0.0064 &  0.0007 \\
BRK      &   202.45 &   8.669 &    1.34 &    2.1 &  12.53 &  0.0257 &  0.0107 \\
INCY     &    97.51 &   8.723 &    0.64 &    2.2 &   9.68 &  0.0095 &  0.0007 \\
MMM      &   222.45 &   8.976 &    0.70 &    1.4 &   9.74 &  0.0334 &  0.0167 \\
WYNN     &   162.67 &   9.496 &    0.76 &    0.2 &   2.17 &  0.0106 &  0.0004 \\
ALXN     &   122.92 &   9.738 &    0.69 &    0.9 &   6.73 &  0.0154 &  0.0013 \\
WPI      &   176.75 &  10.580 &    0.70 &    0.1 &   3.37 &  0.0575 &  0.0093 \\
BA       &   297.60 &  10.630 &    0.97 &    0.2 &   3.31 &  0.0404 &  0.0037 \\
VRTX     &   155.65 &  11.509 &    0.56 &    1.0 &   8.18 &  0.0105 &  0.0009 \\
EQIX     &   433.69 &  11.889 &    0.20 &    2.3 &  12.98 &  0.0015 &  0.0002 \\
FDX      &   246.89 &  12.048 &    0.36 &    0.4 &   7.53 &  0.0445 &  0.0183 \\
ORLY     &   224.90 &  12.416 &    0.43 &    0.8 &   8.09 &  0.0119 &  0.0009 \\
REGN     &   372.48 &  12.480 &    0.30 &    3.4 &  16.62 &  0.0028 &  0.0002 \\
ILMN     &   217.03 &  12.989 &    0.34 &    0.7 &   7.93 &  0.0178 &  0.0033 \\
CHTR     &   344.48 &  13.312 &    0.67 &    0.9 &   8.88 &  0.0060 &  0.0005 \\
IDPH     &   307.01 &  13.479 &    0.54 &    0.8 &   9.68 &  0.0099 &  0.0014 \\
AMZN     &  1097.54 &  14.080 &    1.34 &  160.6 &  31.19 &  0.0117 &  0.2002 \\
GOOG     &  1009.80 &  14.339 &    0.66 &    0.2 &   5.03 &  0.0009 &  0.0001 \\
\bottomrule
\end{tabular}
\captionof{table}{  US stocks (large cap.) used for the empirical analysis of Sec.~\ref{empiricalsec} (see also Figs.~\ref{imm:fit_stoks1}, \ref{imm:fit_stoks2} and \ref{imm:fit_stoks3}). The average spread S has tick units, $V_\textrm{d}$ denotes the average daily traded volume (in shares), $k^{-1}$, $\ell_\ell$ and $\ell_\textrm{r}$ are expressed in $\%$price, and  $\mathcal L$ is shares per unit $\%$price. } \label{tablestocks}
	\end{multicols}
}
	\end{changemargin}

\end{document}